\documentclass[letterpaper,twocolumn,10pt]{article}
\usepackage{usenix}
\usepackage[T1]{fontenc}

\usepackage{graphicx}
\usepackage{booktabs}
\usepackage{amsmath}
\usepackage{amssymb}
\usepackage{hyperref}
\usepackage{cleveref}
\usepackage{subcaption}
\usepackage{multirow}
\usepackage{array}
\usepackage{tikz}
\usepackage{makecell}
\usepackage[table]{xcolor}
\usepackage{colortbl}
\usepackage{tabularx}
\usepackage{algorithm}
\usepackage{algpseudocode}
\usepackage{placeins}

\definecolor{lightgray}{gray}{0.9}

\begin{document}

\date{}

\title{\Large \bf How Well Can LLM Agents Simulate End-User Security and Privacy Attitudes and Behaviors?}

\author{
{\rm Yuxuan Li$^{1*}$}\quad
{\rm Leyang Li$^{2*}$}\quad
{\rm Hao-Ping (Hank) Lee$^{1}$}\quad
{\rm Sauvik Das$^{1}$}\\[0.5em]
{\small $^{1}$School of Computer Science, Carnegie Mellon University}\\
{\small $^{2}$Department of Computer Science and Engineering, University of Notre Dame}\\[0.3em]
{\small $^{*}$Equal contribution}\\[0.2em]
{\small \texttt{yuxuanll@andrew.cmu.edu, lli27@nd.edu, haopingl@cs.cmu.edu, sauvik@cmu.edu}}
}

\maketitle

\begin{abstract}
A growing body of research assumes that large language model (LLM) agents can serve as proxies for how people form attitudes toward and behave in response to security and privacy (S\&P) threats.
If correct, these simulations could offer a scalable way to forecast S\&P risks in products prior to deployment.
We interrogate this assumption using \textsc{SP-ABCBench}, a new benchmark of 30 tests derived from validated S\&P human-subject studies, which measures alignment between simulations and human-subjects studies on a 0-100 ascending scale, where higher scores indicate better alignment across three dimensions: Attitude, Behavior, and Coherence.
Evaluating twelve LLMs, four persona construction strategies, and two prompting methods, we found that there remains substantial room for improvement: all models score between 50 and 64 on average.
Newer, bigger, and smarter models do not reliably do better and sometimes do worse.
Some simulation configurations, however, do yield high alignment: e.g., with scores above 95 for some behavior tests when agents are prompted to apply bounded rationality and weigh privacy costs against perceived benefits.
We release \textsc{SP-ABCBench}~\footnote{\url{https://anonymous.4open.science/r/ABCBench-B808/README.md}} to enable reproducible evaluation as methods improve.
\end{abstract}

\section{Introduction}
\label{sec:intro}
When endowed with \textit{personas}\footnote{These personas can range from crude demographic variables to rich personal narratives.} designed to represent people, LLM agents have been used to simulate social reasoning, collective decision-making, behavioral diversity, and even adversarial misuse of online platforms~\cite{park2023generative, park2022social, wang2024farsight, li2025makes, binz2025foundation}. As a result, in domains such as online safety \cite{zhouhaicosystem}, privacy and security \cite{zhou2025safeagent}, and red teaming \cite{asadi2025personas}, these simulations are increasingly being used to anticipate user reactions, identify risks, and test system robustness.
As one team put it: ``LLM-driven personas can serve as crash dummies for experimenting on security and privacy scenarios''~\cite{asaditoward}.

But can they? An assumption of these efforts is that LLM-powered generative agents are plausible proxies for real users.
Yet, there has been little validation of how well these simulations reproduce empirically observed human behavior in S\&P contexts.
S\&P decision-making differs in important ways from many domains in which LLM agents have been evaluated.
S\&P concerns are often secondary to users' primary goals~\cite{whitten1999johnny, adams1999users}, preventive rather than immediately rewarding~\cite{anderson2001information, campbell2003economic}, and oriented toward abstract or invisible threats~\cite{friedman2002informed, wash2010folk, acquisti2005privacy, das2022security}.
As a result, users' S\&P decisions are rarely optimal or perfectly rational; instead, users commonly rely on heuristic-based cost–benefit trade-offs and contextual cues as captured by theoretical frameworks such as, e.g., Privacy Calculus, Bounded Rationality, the Security and Privacy Acceptance Framework (SPAF), and Contextual Integrity ~\cite{culnan1999information, dinev2006extended, simon1955behavioral, das2022security, nissenbaum2004privacy}.
These characteristics raise fundamental questions about whether existing models, persona construction techniques, and prompting strategies can capture how people reason coherently about, form attitudes toward, and act upon S\&P risks.
In this work, we ask:
\begin{itemize}
\item \textbf{RQ1 (Models):} How well can current LLMs reproduce population-level S\&P patterns? Do model scale, vintage, or reasoning capability predict simulation quality?
\item \textbf{RQ2 (Persona Construction):} How do different persona construction strategies affect alignment between simulated outcomes and empirical human data?
\item \textbf{RQ3 (Prompting Strategy):} To what extent can prompting strategies informed by S\&P-theory improve LLMs’ ability to reproduce population-level S\&P patterns?
\end{itemize}

To address these questions, we introduce \textsc{SP-ABCBench} (Security and Privacy Attitude--Behavior--Coherence Benchmark), an extensible benchmark grounded in validated empirical S\&P research.
\textsc{SP-ABCBench} comprises 30 empirical tests derived from 15 foundational human-subject studies on S\&P attitudes and behaviors~\cite{malhotra2004_IUIPC, egelman2015_SeBIS, faklaris2019_SA6, buchanan2007_OPC1, acquisti2015_PHB, Tsai2011premium_spending_for_privacy, Singer1992attendance_rate_without_vs_with_assurance, Acquisti2013ratio_keep_switch_anonomous_card, Leslie2010casual_vs_professional_interface, Acquisti2012disclosure_rate_high_vs_missing_norm, Stutzman2010switch_visibility_with_vs_without_unintended_sharing, Goldfarb2011setting_visit_rate_with_generic_vs_personalized_ads, Hoofnagle2014disclosure_rate_with_vs_without_privacy_policy, Xu2009LBS_usage_with_vs_without_gov_regulation, Brandimarte2013disclosure_rate_high_vs_low_user_control}.
Each test captures a measurable, population-level effect under experimentally validated conditions.
The benchmark is organized along three theoretically grounded dimensions.
\emph{Attitude} captures beliefs, concerns, and intentions related to S\&P.
\emph{Behavior} captures observable actions in realistic S\&P scenarios.
\emph{Coherence} captures relationships among S\&P-related constructs, such as correlations between psychometric scales, demographic differences, or structural paths in causal models.
Together, these dimensions span how people think about, feel about, and act upon S\&P risks.
We also introduce a simulation quality metric that quantifies alignment between simulated and empirical results on a scale from 0 (no alignment) to 100 (perfect alignment).

Our evaluation covers twelve models from four major model families, including both open- and closed-source systems spanning a range of scales and reasoning capabilities (RQ1).
We examine four persona construction strategies drawn from prior work, including two general-purpose methods and two that require adaptation to S\&P-specific contexts (RQ2).
In addition, we develop a theory-informed prompting strategy grounded in Privacy Calculus and Bounded Rationality~\cite{culnan1999information, dinev2006extended, simon1955behavioral, wei2022chain} for S\&P decision-making carried out under limited attention and imperfect reasoning (RQ3).

Our findings suggest that \textbf{current models exhibit only moderate population-level alignment on average.}
Across all models, average simulation quality scores across all tests on \textsc{SP-ABCBench} range from 50 to 64 out of 100.
Counterintuitively, bigger, newer, and smarter models do not always score higher and sometimes score worse.
However, there are specific simulation configurations that perform well for specific tests (with scores above 95).
Second, \textbf{persona construction strategies vary in effectiveness across Attitude, Behavior, and Coherence tests.}
For example, Scenario-Primed Personas substantially improve performance on Behavior tests, while their effectiveness for Attitude and Coherence tests varies across models.
Finally, \textbf{theory-informed prompting improves simulation quality scores across all persona construction strategies.}
Our contributions include:
\begin{enumerate}
\item We systematically evaluate how well current LLMs can simulate population-level S\&P attitudes and behaviors.
\item We introduce \textsc{SP-ABCBench}, a benchmark grounded in human-subject studies that enables evaluation of LLM agent simulations in S\&P contexts.
\item We empirically characterize the effects of model choice, persona construction, and theory-informed prompting, offering practical guidance when using LLM simulations for S\&P risk assessment.
\end{enumerate}
\section{Related Work}
\label{sec:related}

\subsection{Envisioning Security \& Privacy Risks}

Security and privacy (S\&P) practitioners use a variety of tools to assess and anticipate risks during product development.
Common approaches include checklists~\cite{deng2025supporting, bogucka2024co}, card-based exercises~\cite{linddun_go_2026}, government frameworks~\cite{ai2023artificial}, privacy impact assessments~\cite{clarke2009privacy, bogucka2024ai}, and threat modeling~\cite{shostack2014threat, kohnfelder1999threats}.
These methods provide structured guidance but depend heavily on the knowledge, assumptions, and experience of the assessor, often resulting in blind spots or inconsistent coverage~\cite{emami2021privacy, wong2019bringing, warford2022sok, thompson2024there, lee2024i}.
But they are also labor-intensive, difficult to scale, and require frequent updates as technologies and threat landscapes evolve, prompting work in responsible AI to explore the use of large language models (LLMs) to support risk envisioning ~\cite{buccinca2023aha, wang2024farsight, herdel2024exploregen}.

Within S\&P, emerging approaches use LLMs to identify privacy risks or to generate threat models from system descriptions~\cite{lee2025privy, mollaeefar2024pillar}.
Concurrent work extends LLM agent simulation to S\&P risk discovery, including red-teaming frameworks that alternate attack and defense agents to surface vulnerabilities~\cite{zhang2025searching}, safety simulators that stress-test agent--user interactions at scale~\cite{zhou2025safeagent, zhouhaicosystem}, and LLM-driven personas proposed as ``crash dummies'' for security testing~\cite{asaditoward}.
These efforts focus on using LLMs to \emph{generate} or \emph{provoke} risky behaviors; our work complements this work by empirically measuring how well suited LLMs are to this task by assessing whether LLM agents effectively reproduce population-level S\&P patterns established in empirical human-subject studies.

\subsection{LLM Agent Simulation}

LLMs have increasingly been deployed as simulated agents to model human decision-making and social interaction.
Their broad world knowledge and ability to condition on detailed persona descriptions make them a natural substrate for behavioral simulation~\cite{argyle2023out, binz2025foundation}.
Early work showed that agents equipped with memory, planning, and communication capabilities can coordinate activities, form social relationships, and exhibit emergent collective behavior~\cite{park2023generative, binz2025foundation, chen2024agentverse, zhou2025sotopia, yang2024oasis}.
Subsequent research has extended agent-based simulation to domains such as healthcare workflows, education, economic markets, and policy analysis, motivated in part by the ability to run counterfactual experiments that would be costly or infeasible with human participants~\cite{li2024agent, zhang2025simulating, karten2025llm, ji2024srap} --- e.g., epidemic response planning, misinformation spread, and urban systems~\cite{li2025makes, williams2023epidemic, liu2025stepwise, yan2024opencity}.

Prior agent-simulation work mainly focuses on domains with observable outcomes, immediate feedback, or clearly defined objectives.
Nevertheless, simulating S\&P behaviors entails key differences: such behaviors are often 1) secondary to users’ primary tasks~\cite{whitten1999johnny, adams1999users}, 2) preventive rather than reactive~\cite{anderson2001information, campbell2003economic}, and 3) oriented toward abstract or probabilistic harms~\cite{friedman2002informed, wash2010folk, acquisti2005privacy, das2022security}.
As a result, it remains unclear whether LLM agents can faithfully reproduce empirically documented population-level patterns in S\&P attitudes, behavior, or coherence.
Our work addresses this gap by systematically validating LLM agent simulations against established S\&P findings from prior work.

\subsection{S\&P Attitudes, Behaviors, and Coherence}

Empirical S\&P research characterizes how people reason about risks, form attitudes, and act in real-world settings.
Validated psychometric instruments quantify privacy concern and security posture.
IUIPC models online privacy concern and its relationships with trust, risk, and behavioral intention~\cite{malhotra2004_IUIPC}.
SeBIS and SA-6 capture intentions and attitudes toward common security practices~\cite{egelman2015_SeBIS, faklaris2019_SA6}, while OPC provides brief Internet-administered measures of privacy concern and protective tendencies~\cite{buchanan2007_OPC1}.
Together, these instruments establish reliable constructs, subscale structure, and expected correlations that support population-level distributional and structural comparisons.

Behavioral studies further document how individuals make S\&P decisions in laboratory, field, and in-the-wild contexts.
Findings include willingness to pay for protection~\cite{Tsai2011premium_spending_for_privacy, Hann01102007, Acquisti2013ratio_keep_switch_anonomous_card}, sensitivity to consent architectures such as defaults and banners~\cite{Gross2005InformationRA, Liu2011AnalyzingFP, Stutzman2013SilentLT, Utz2019UninformedCS}, and responses to warnings and permission prompts~\cite{Felt2012AndroidPU, Utz2019UninformedCS}.
Across settings, prior work emphasizes uncertainty, context dependence, and the malleability of preferences under interface and institutional cues~\cite{acquisti2015_PHB}.

Additionally, complementary theoretical frameworks explain these patterns.
The Privacy Calculus model frames decisions as risk--benefit trade-offs moderated by trust and uncertainty~\cite{culnan1999information, dinev2006extended}; bounded rationality accounts for systematic deviations from optimal behavior due to cognitive constraints~\cite{simon1955behavioral}.
We draw on these theories to define benchmark dimensions (Section~\ref{sec:benchmark}) and to inform our theory-driven prompting (Section~\ref{subsec:theory_informed_prompting_experiment}) of LLM agents.
\section{\textsc{SP-ABCBench}}
\label{sec:benchmark}

We construct \textsc{SP-ABCBench}, a benchmark of 30 empirical tests derived from 15 foundational human-subject studies in S\&P research.
The benchmark evaluates alignment with published population-level S\&P attitudes and behaviors rather than individual-level prediction.
\textsc{SP-ABCBench} is designed as an extensible foundation; we expect future work to expand its coverage across domains, cultures, and interaction modalities.

\subsection{Benchmark Dimensions}
\label{sec:dimensions}

How people \emph{think} about S\&P risks, how they \emph{act} in response, and whether their thoughts and actions are \emph{internally consistent} are related but empirically distinct phenomena~\cite{acquisti2015_PHB}.
A useful simulation must capture all three: reproducing survey responses alone is insufficient if behavioral effects are missed, and matching both is insufficient if documented inter-construct relationships break down.
\textsc{SP-ABCBench} accordingly organizes its 30 tests along three dimensions:
\textbf{Attitude (4 tests)} captures the distribution of responses to validated S\&P survey instruments---e.g., how concerned a representative sample reports being about online data collection, or how strongly they endorse proactive security behaviors.
\textbf{Behavior (11 tests)}
captures observable actions in experimentally controlled S\&P scenarios---e.g., whether users disclose more on a casual quiz interface than on a formal institutional survey~\cite{Leslie2010casual_vs_professional_interface}.
\textbf{Coherence (15 tests)}
captures whether documented \emph{relationships} among S\&P constructs are preserved in simulation: correlations between psychometric scales (e.g., security attitudes and behavior intentions), structural paths in causal models (e.g., privacy concern $\rightarrow$ perceived risk $\rightarrow$ behavioral intention), and demographic group differences (e.g., older adults scoring higher on security attitudes).
Table~\ref{tab:tests} summarizes all benchmark components.

\begin{table}[t]
\centering
\caption{Empirical tests in \textsc{SP-ABCBench}, grouped by dimension. \textit{SEM: structural equation model; LBS: location-based services}.}
\label{tab:tests}
\footnotesize
\setlength{\tabcolsep}{4pt}
\begin{tabular}{p{0.40\columnwidth} p{0.52\columnwidth}}
\toprule
\textbf{Test} & \textbf{Key Measure} \\
\midrule
\rowcolor{lightgray}
\multicolumn{2}{l}{\textit{Attitude (4)}} \\
IUIPC~\cite{malhotra2004_IUIPC}
  & Subscale score distributions \\
SeBIS~\cite{egelman2015_SeBIS}
  & Item-level response distributions \\
SA-6~\cite{faklaris2019_SA6}
  & Security attitude distribution \\
OPC~\cite{buchanan2007_OPC1,joinson2010_OPC2}
  & Subscale score distributions \\
\midrule
\rowcolor{lightgray}
\multicolumn{2}{l}{\textit{Behavior (11)}} \\
Privacy Premium~\cite{Tsai2011premium_spending_for_privacy}
  & Willingness to pay for privacy \\
Confidentiality Assurance~\cite{Singer1992attendance_rate_without_vs_with_assurance}
  & Survey participation ($\pm$ assurance) \\
Gift Card Anonymity~\cite{Acquisti2013ratio_keep_switch_anonomous_card}
  & Anonymous vs.\ trackable choice \\
Interface Formality~\cite{Leslie2010casual_vs_professional_interface}
  & Disclosure under casual vs.\ professional UI \\
Normative Info (2)~\cite{Acquisti2012disclosure_rate_high_vs_missing_norm}
  & Disclosure under peer norms \\
Unintended Audience~\cite{Stutzman2010switch_visibility_with_vs_without_unintended_sharing}
  & Restriction after unintended exposure \\
Personalized Ads~\cite{Goldfarb2011setting_visit_rate_with_generic_vs_personalized_ads}
  & Settings visits post-exposure \\
Policy Link~\cite{Hoofnagle2014disclosure_rate_with_vs_without_privacy_policy}
  & Disclosure ($\pm$ privacy policy link) \\
Regulatory Assurance~\cite{Xu2009LBS_usage_with_vs_without_gov_regulation}
  & LBS adoption ($\pm$ regulation) \\
Control Paradox~\cite{Brandimarte2013disclosure_rate_high_vs_low_user_control}
  & Disclosure under explicit control \\
\midrule
\rowcolor{lightgray}
\multicolumn{2}{l}{\textit{Coherence (15)}} \\
Privacy Structural Paths~\cite{malhotra2004_IUIPC}
  & SEM: concern $\rightarrow$ trust/risk $\rightarrow$ intention \\
Security Intentions (5)~\cite{egelman2015_SeBIS}
  & Correlations with psychometric scales \\
SA-6 Convergent (4)~\cite{faklaris2019_SA6}
  & Correlations with SeBIS, impulsivity, self-efficacy \\
SA-6 Demographics (4)~\cite{faklaris2019_SA6}
  & Group differences by age, gender, education, income \\
OPC Intercorrelations~\cite{buchanan2007_OPC1,joinson2010_OPC2}
  & Inter-subscale correlations \\
\bottomrule
\end{tabular}
\end{table}

\subsection{Source Study Selection}
\label{sec:source_selection}

We selected source studies that (1)~report a validated, population-level finding---a distributional statistic, effect size, or structural relationship---usable as a quantitative evaluation target; (2)~have experimental conditions reproducible through text- or image-based prompts, excluding studies that require physical environments, sensory manipulation, or longitudinal logs; and (3)~are widely cited or replicated.
These criteria yielded two complementary bodies of work.

\subsubsection{Psychometric Instruments (Attitude \& Coherence)}
\label{sec:psychometric}

We selected four widely validated instruments measuring S\&P attitudes: Internet Users' Information Privacy Concerns (IUIPC)~\cite{malhotra2004_IUIPC}, Security Behavior Intentions Scale (SeBIS)~\cite{egelman2015_SeBIS}, Security Attitudes (SA-6)~\cite{faklaris2019_SA6}, and Online Privacy Concern (OPC)~\cite{buchanan2007_OPC1}.
Each contributes to both Attitude and Coherence tests.

For \emph{Attitude}, we compare simulated Likert-scale response distributions to the published distributions.
For example, the IUIPC study reports means and standard deviations for privacy-concern subscales along with trust, perceived risk, and behavioral intention; an aligned simulation should produce aggregate score distributions whose means deviate only slightly (relative to the pooled standard deviation) from these published values.

For \emph{Coherence}, we evaluate whether documented inter-construct relationships are preserved.
IUIPC reports SEM coefficients linking privacy concern, trust, risk, and behavioral intention.
SeBIS documents correlations with five companion scales: risk attitude (DoSpeRT), decision-making style (GDMS), need for cognition (NFC), impulsivity (BIS), and future orientation (CFC)~\cite{blais2006DoSpeRT, scott1995GDMS, cacioppo1984NFC, Stanford2009385BIS, joireman2012CFC}.
SA-6 reports convergent correlations with SeBIS and self-efficacy measures~\cite{schwarzer1995GSE, sherer1982SSE}, as well as group differences by age, gender, education, and income.
OPC reports intercorrelations among its subscales~\cite{joinson2010_OPC2}.
Together, these instruments yield 4 Attitude and 15 Coherence tests.

\subsubsection{Behavioral Experiments (Behavior)}
\label{sec:behavioral}

We drew from Acquisti et al.'s review on digital privacy behavior~\cite{acquisti2015_PHB}, selecting 11 experiments with measurable behavioral effects that can be operationalized through text- or image-based scenarios.
Each experiment manipulates a contextual factor and measures how it shifts an S\&P-relevant behavior.

Two examples illustrate the range.
In the \emph{Control Paradox} test, Brandimarte et al.~\cite{Brandimarte2013disclosure_rate_high_vs_low_user_control} found that giving participants explicit control over which survey answers would be published paradoxically \emph{increased} disclosure of sensitive information; our benchmark tests whether simulated agents exhibit the same pattern.
In the \emph{Gift Card Anonymity} test~\cite{Acquisti2013ratio_keep_switch_anonomous_card}, participants endowed with a \$10 anonymous gift card tend to keep it when offered a \$12 trackable alternative, while those holding the trackable card also tend to keep theirs---indicating that the endowment, not absolute value, drives the choice.

Across the 11 experiments, tests span disclosure decisions, adoption choices, and behavioral responses to contextual cues including peer norms, confidentiality assurances, interface design, and regulatory signals.

\subsection{Operationalizing Tests for LLM Agents}
\label{sec:operationalization}

Each test is administered by prompting an \emph{LLM agent}---a language model endowed with a persona and presented with a stimulus---to respond as a simulated study participant.
Personas are constructed from demographic attributes (age, sex, education, income) sampled to match the original study populations; Section~\ref{sec:experiments} describes persona construction in detail.
Each agent produces a single response; responses are aggregated across the simulated population for evaluation.

\paragraph{Survey simulations.}
Agents complete the IUIPC, SeBIS, SA-6, and OPC questionnaires, plus the companion instruments needed for Coherence evaluation.
Responses are scored using the same constructs and aggregation procedures as the original studies.
For example, to evaluate SeBIS we prompt each agent to answer all 16 SeBIS Likert items and the five companion scales (DoSpeRT, GDMS, NFC, BIS, CFC), then apply the published scoring rules: reverse-coding designated items, averaging into the four SeBIS subscales, and computing companion-scale totals.
Survey simulations yield 4 Attitude and 15 Coherence tests.

\paragraph{Scenario simulations.}
Agents read a textual or visual description of an S\&P situation and choose among predefined actions.
Each behavioral experiment is operationalized as a pair of scenarios that differ in a single contextual manipulation, mirroring the original experimental conditions.
For example, the \emph{Confidentiality Assurance} test~\cite{Singer1992attendance_rate_without_vs_with_assurance} assigns agents to one of two invitation prompts---a plain survey invitation or the same invitation augmented with a detailed confidentiality statement---and each agent returns a binary decision (attend or decline).
We aggregate decisions per condition and score by whether the simulation reproduces the published direction and magnitude.
One exception is the Normative Information test~\cite{Acquisti2012disclosure_rate_high_vs_missing_norm}, which measures how people's willingness to disclose sensitive information is impacted by social norms.
It uses three scenarios (no norm, low-admission norm, high-admission norm) to evaluate two contrasts (high vs.\ low; high vs.\ absent), yielding two Behavior items from a single experiment.
In total, scenario simulations comprise 21 scenarios producing 11 Behavior tests.

\subsection{Quantifying Simulation Quality}
\label{sec:scoring}

Our tests span fundamentally different statistical quantities---distributional distances, effect-size ratios, and correlation errors---so no single off-the-shelf metric applies.
We define a per-dimension scoring function that maps outputs to a unified 0--100 scale, where 0 indicates no alignment (e.g., a simulated effect in the wrong direction) and 100 indicates perfect alignment (e.g., distributions indistinguishable from published ones).
Each function includes a tolerance parameter $\tau$ that controls how steeply scores decline with deviation from the published target; we set $\tau$ values to reflect typical effect and coefficient magnitudes in the source studies.

\paragraph{Attitude.}
We compute Cohen's $d$~\cite{cohen1962statistical} between the simulated and published score distributions and average across subscales ($d < 0.2$ is conventionally negligible).
Given $d$ and tolerance $\tau = 2.0$:
\begin{equation}
s = \text{clip}_{[0,1]}\left(1 - \frac{|d|}{\tau}\right)
\end{equation}
For instance, if simulated IUIPC ``Collection'' scores have a mean of 5.8 versus the published 5.5 at similar variance, $d \approx 0.25$ yields $s = 87.5$; a simulated mean of 3.0 would push $s$ toward zero.

\paragraph{Behavior.}
We compute a ratio of the target behavior rate across conditions (e.g., disclosure rate on a casual interface divided by the rate on a professional interface).
Ratios above 1 indicate the simulation reproduces the published effect direction; ratios at or below 1 indicate a reversal.
Reversals are penalized linearly; large ratios are log-compressed to limit outlier influence.
Given ratio $r$ and cap $R_{\max} = 100$:
\begin{equation}
s =
\begin{cases}
0, & r \leq 0 \\
0.5\, r, & 0 < r \leq 1 \\
0.5 + 0.5\;\dfrac{\ln(\min\{r,\, R_{\max}\})}{\ln R_{\max}}, & r > 1
\end{cases}
\end{equation}
For the Interface Formality test, if 70\% of agents disclose on the casual interface versus 50\% on the professional one, $r = 1.4$ yields a score above 50.
If the simulation instead produces $r = 0.8$ (wrong direction), the score drops to 40.

\paragraph{Coherence.}
For correlation-based tests, we compute the mean absolute error between simulated and published coefficients.
For demographic group differences, we compute Welch's $t$-statistics~\cite{welch1938significance} in the published contrast direction.
Both use:
\begin{equation}
s = \text{clip}_{[0,1]}\left(1 - \frac{|x - x^*|}{\tau}\right)
\end{equation}
with $\tau = 0.5$ for correlations (so that deviations on the order of a medium-sized coefficient are penalized) and $\tau = 2.0$ for $t$-statistics (roughly matching conventional significance thresholds).
For example, if the published correlation between SA-6 and SeBIS is $\rho^* = 0.45$ and the simulation yields $\rho = 0.30$, then $s = 1 - 0.15/0.5 = 70$; a negative simulated correlation would score near zero.

\paragraph{Aggregation and robustness.}
Scores are clipped to $[0,1]$, scaled to 0--100, and averaged within each dimension and overall.
To verify that our findings are insensitive to specific $\tau$ choices, we varied each tolerance by $\pm 25\%$ and measured ranking stability via Kendall's rank correlation~\cite{kendall1938new}: all rank correlations exceed 0.90 and the mean absolute rank change is under 1.5 positions across all persona strategies (Appendix~\ref{app:tau_sensitivity}).
\section{Experiments}
\label{sec:experiments}

Using \textsc{SP-ABCBench}, we evaluate how well LLM agent simulations reproduce population-level S\&P patterns reported in human-subject studies.
Our experiments are organized around three research questions:
\textbf{RQ1 (Models)} asks whether model scale, vintage, or reasoning capability predicts simulation quality;
\textbf{RQ2 (Persona Construction)} asks how different strategies for endowing agents with identity information affect alignment with empirical data; and
\textbf{RQ3 (Prompting Strategy)} asks whether prompting grounded in S\&P theory improves simulation fidelity.

\begin{figure*}[t]
\centering
\includegraphics[width=0.78\textwidth]{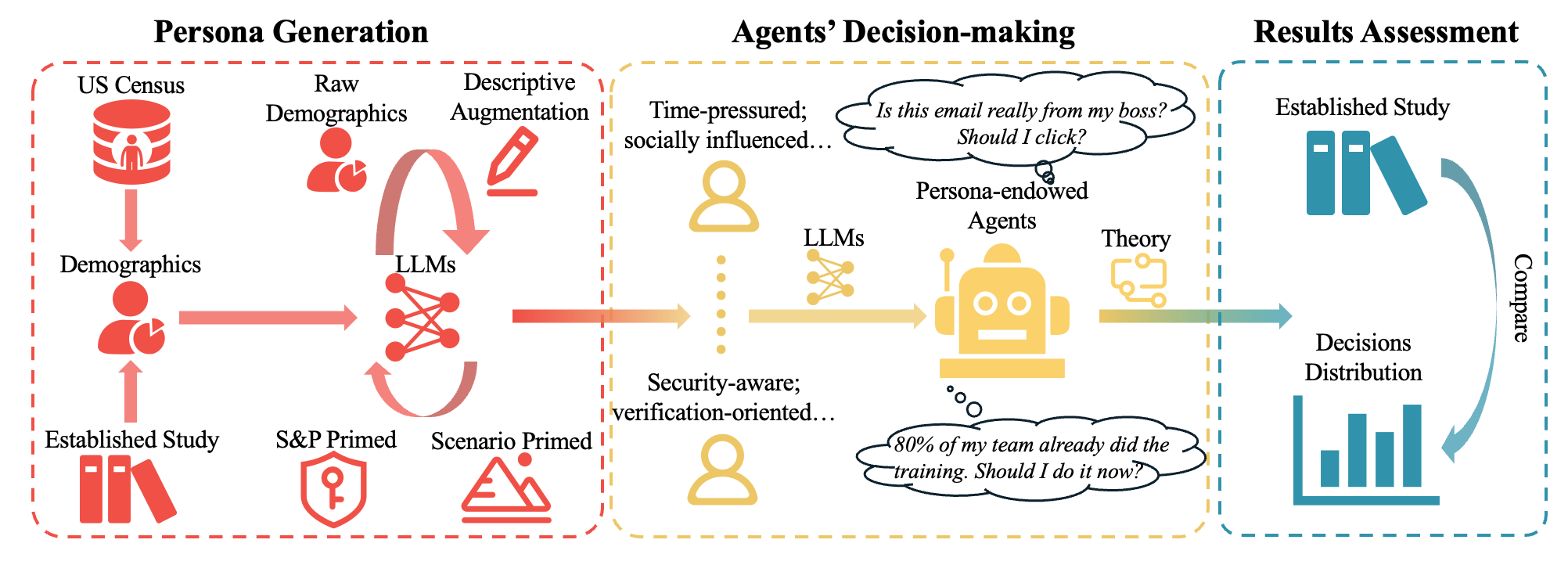}
\caption{Experimental pipeline for population-level S\&P simulation.
\textbf{Left:} We generate synthetic participants by sampling U.S. Census--based demographics and constructing personas using four strategies of increasing specificity (RQ2).
\textbf{Center:} Persona-endowed LLM agents (RQ1) complete S\&P survey instruments or decision tasks, optionally guided by theory-informed prompting (RQ3).
\textbf{Right:} Aggregated agent responses are compared against published human-subject results to evaluate alignment across Attitude, Behavior, and Coherence.}
\label{fig:pipeline}
\end{figure*}

To isolate each factor, we hold the demographic composition and task structure constant while varying the model (RQ1), persona construction strategy (RQ2), and prompting method (RQ3).
Figure~\ref{fig:pipeline} illustrates the full pipeline.

\subsection{Experimental Setup}

For each empirical test in \textsc{SP-ABCBench}, we:
(i)~construct a synthetic population of agents whose demographic distribution matches that of the original study (or a U.S. Census--representative sample when such information is unavailable);
(ii)~prompt each agent to act as a study participant by completing either a survey instrument or a decision-making task;
(iii)~aggregate agent responses and compute simulation quality scores defined in Section~\ref{sec:benchmark}; and
(iv)~repeat the evaluation across multiple LLMs (RQ1), persona construction strategies (RQ2), and prompting conditions (RQ3).

\subsubsection{Models (RQ1)}

To address RQ1, we evaluate twelve large language models spanning four families:
GPT-4.1-Nano, GPT-4.1-Mini, GPT-4.1,
GPT-5-Nano (Minimal Reasoning), GPT-5-Mini (Minimal Reasoning),
GPT-5 (Minimal Reasoning), GPT-5 (Medium Reasoning),
Gemini-2.5-Flash-Lite, Gemini-2.5-Flash,
Gemini-3.0-Flash,
Qwen3-Next-80B, and Llama-4-Maverick.
These models are not exhaustive, but are representative of a wide gamut of model types including both closed- and open-source models that vary in parameter scale, release date, and reasoning configuration.
We access OpenAI models via the OpenAI API~\cite{openai_docs}, Gemini models via the Google API~\cite{google_gemini_docs}, and open-source models via the Together AI platform~\cite{together_ai}.

\subsubsection{Persona Construction Strategies (RQ2)}
\label{subsubsec:PCS}

To address RQ2, we evaluate four persona construction strategies that reflect common practices in prior work on LLM agent simulation~\cite{argyle2023out, ge2024scaling, joshi-etal-2025-improving, li2025actions, liu2025mosaic}.
Each strategy specifies how demographic attributes are represented and whether additional contextual information is introduced.
While not exhaustive, these strategies cover both widely used baselines and approaches we hypothesize to be relevant for S\&P contexts.
Moreover, we expect that other promising persona construction strategies may also be benchmarked against, as they emerge, in future work \textsc{SP-ABCBench}.

\textbf{Strategy~I: Demographic.}
Agents receive only basic demographic attributes expressed as labeled fields without elaboration---e.g., \textit{``Sex: Male. Age: 34. Education: Bachelor's degree. Income: \$55{,}000.''}
This minimal baseline, common in survey-simulation studies~\cite{argyle2023out}, tests whether coarse population conditioning alone is sufficient.

\textbf{Strategy~II: Raw Persona.}
Agents receive demographic attributes along with a descriptive persona generated by an LLM.
The generation prompt adds plausible personal details---name, personality traits, hobbies, social context, and life experiences---while avoiding explicit references to S\&P attitudes.
For example: \textit{``Your name is Marcus, a 34-year-old warehouse supervisor who coaches his daughter's soccer team on weekends\ldots''}
This approach follows the biographical augmentation strategy used in prior work~\cite{ge2024scaling}.

\textbf{Strategy~III: S\&P-Primed Persona.}
In addition to demographics, agents are endowed with explicit descriptions of their general S\&P orientations.
The persona generation prompt asks the model to characterize how the individual typically perceives, values, and manages S\&P in everyday life, without reference to a specific experimental task---e.g., \textit{``\ldots Marcus values his privacy but rarely thinks about online security day-to-day\ldots''}
This strategy draws on prior work showing that domain-specific priming improves simulations~\cite{joshi-etal-2025-improving}.

\textbf{Strategy~IV: Scenario-Primed Persona.}
Agents receive persona descriptions tailored to the specific construct examined in each empirical test.
For example, when simulating IUIPC, the persona includes beliefs related to perceived risk, trust, and control over personal information; for behavioral experiments, personas emphasize situational traits relevant to the decision context---e.g., \textit{``\ldots Marcus accepts app permissions quickly when the app seems useful, even if the data request feels excessive\ldots''}
This strategy, analogous to task-conditioned persona generation in social simulation~\cite{li2025actions, liu2025mosaic}, aligns persona content with the theoretical focus of each study.

\subsubsection{Theory-Informed Prompting (RQ3)}
\label{subsec:theory_informed_prompting_experiment}

To address RQ3, we introduce a theory-informed prompting condition grounded in Privacy Calculus and Bounded Rationality~\cite{culnan1999information, dinev2006extended, simon1955behavioral}.
This prompt encourages agents to reason in ways consistent with everyday decision-making, emphasizing trade-offs among convenience, effort, trust, and perceived risk rather than fully rational optimization.
The full prompt text is provided in Appendix~\ref{app:prompts}.
We note that these are not the only possible prompting approaches; for example, other theoretical frameworks such as Contextual Integrity~\cite{nissenbaum2004privacy} and SPAF~\cite{das2022security} could also be used. 
We employ the ``theory-informed'' approach here as a representative of a broader class of prompting strategies that aim to guide agents to act in accordance with usable S\&P theory.

\subsubsection{Demographic Sample Construction}

For each source study, we construct a synthetic sample that mirrors reported participant demographics.
We sample individuals from the U.S. Census Public Use Microdata Sample (PUMS)~\cite{uscensus2025}, using four key attributes reported across all studies: sex, age, education level, and income.

We take a two-pronged approach depending on the availability of demographic data in the source studies.
For studies that report participant demographic distributions---IUIPC~\cite{malhotra2004_IUIPC}, SeBIS~\cite{egelman2015_SeBIS}, and SA-6~\cite{faklaris2019_SA6}---we apply stratified sampling to match those distributions directly.
For studies that do not report detailed demographics, we use the best available proxy: for OPC~\cite{buchanan2007_OPC1}, we use distributions from a replication study~\cite{joinson2010_OPC2}; for behavioral experiments derived from Acquisti et al.~\cite{acquisti2015_PHB}, we sample a U.S. internet user population based on demographic statistics reported by Pew Research~\cite{pew_internet_broadband_2025}.
In all cases, the synthetic sample size matches the number of participants reported in the original study.

\subsubsection{Procedure}

Unless otherwise specified, we evaluate all models (RQ1) across all four persona construction strategies (RQ2) on every empirical test in \textsc{SP-ABCBench} using a standard, non-theory-informed prompt.
To isolate the effect of theory-informed prompting (RQ3), we additionally evaluate GPT-5 (Minimal Reasoning) --- selected as a recent, widely accessible model with near-median baseline performance --- under both standard and theory-informed prompting conditions across all persona strategies.

We set the temperature to 1.0 during persona generation to encourage diversity and to 0.3 during task inference to promote response stability.
For each model--persona--test configuration, we collect one response per synthetic agent.
All responses are aggregated and converted into simulation quality scores using the procedures defined in Section~\ref{sec:benchmark}.
We do not perform inferential statistical tests on simulation quality scores.
With temperature set to 0.3, outputs for a given configuration (model, persona, prompting method, test) exhibit minimal run-to-run variance---meaning we could arbitrarily achieve statistical significance by running more simulations.
Instead, following standard LLM benchmark practice, we report observed scores directly as the actual performance differences~\cite{yue2024mmmu, yu2025youthsafe, center2026benchmark}.

\subsubsection{Qualitative Exploration of Agent Rationales}
\label{subsec:qual_methods}
To understand \textit{why} agents produced particular responses, we examined the rationales agents' produced for their outputs across simulation configurations.
We randomly sampled agent outputs stratified across high-scoring (SQ $\geq$ 80) and low-scoring (SQ $<$ 50) configurations for each test type.
Two members of the research team reviewed the sampled traces using open coding~\cite{glaser2017discovery}, identifying representative examples and discussing to reach consensus on key terms indicative of specific reasoning patterns (e.g., mentions of practical constraints, references to contextual cues, application of blanket heuristics).
We then counted occurrences of these key terms across all agents within targeted simulation configurations.
These qualitative findings are intended to aid interpretation of quantitative results and should not be treated as definitive causal explanations for agent behavior~\cite{li2025actions}.

\section{Results}
\label{sec:results}

We found that LLM agent simulations achieved only moderate alignment with the S\&P patterns reported in prior empirical studies: the average simulation quality (SQ) across all tests in \textsc{SP-ABCBench} ranged from 50 to 64 for the models we evaluated.
However, there were specific simulation configurations (i.e., combinations of models, persona construction strategies, and prompting methods) that did achieve high overall SQ on individual tests (26.94\% scoring 80 or above, 13.33\% scoring 90 or above).
Counterintuitively, smaller models often outperformed larger ones, and extended reasoning did not help.
Persona construction strategies exhibited dimension-dependent trade-offs: e.g., Scenario-primed personas improved SQ for Behavior tests but degraded SQ for Attitude tests.
Theory-informed prompting improved SQ for Behavior and Coherence tests but slightly degraded Attitude tests, with the largest gains where baseline performance was weakest.

\begin{figure*}[t]
\centering
\includegraphics[width=0.8\textwidth]{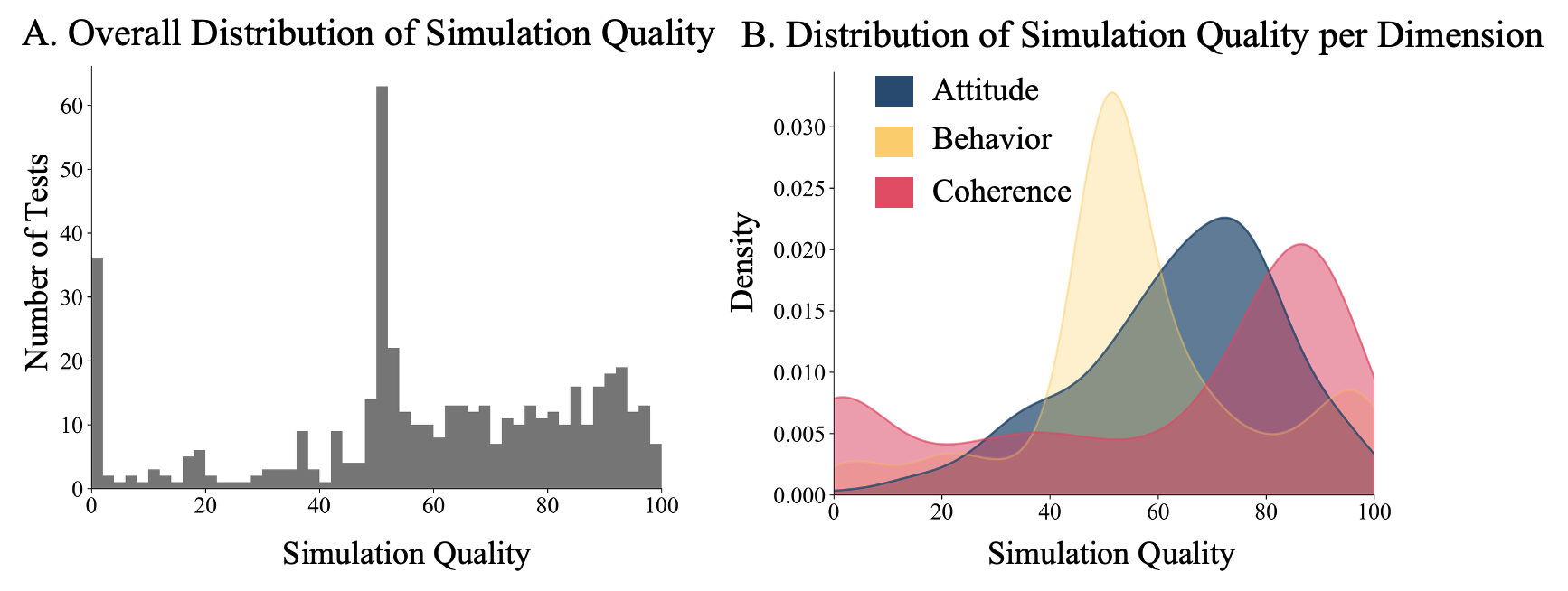}
\caption{Distribution of simulation quality scores. (A) Overall distribution across all tests using demographic attributes only (N = 360, mean = 58.92, SD = 27.38). (B) Density plots by test type show distinct patterns: Attitude tests cluster tightly around 65--75, Behavior tests show lower and more variable scores, and Coherence tests exhibit a bimodal distribution.}
\label{fig:sq-distribution}
\end{figure*}

\subsection{Understanding Simulation Quality}
\label{sec:sq-understanding}

Our simulation quality (SQ) metric evaluated the alignment of simulation outputs with documented \emph{population-level} empirical results from prior work, rather than alignment between any one LLM agent and any one user.

\subsubsection{Score Distributions Across Test Dimensions}

Figure~\ref{fig:sq-distribution} summarizes score distributions under Strategy~I (Demographic).
Scores spanned the full 0--100 range with a mean of 58.92 (SD = 27.38), indicating that simulations aligned well with some tests in \textsc{SP-ABCBench} yet deviated sharply on others.

The three benchmark dimensions showed systematic differences.
\textbf{Attitude} tests were comparatively higher-scoring, on average, and exhibited lower variance (N = 192, mean = 65.90, SD = 18.76).
\textbf{Behavior} tests had the lowest average SQ (N = 528, mean = 56.14, SD = 21.61), as well as the lowest median (51.51).
\textbf{Coherence} tests showed a bimodal distribution (N = 720, mean = 59.21, SD = 34.40), producing a high median (76.45) despite a lower mean.

These patterns likely reflect task structure: Attitude tests require only Likert-scale responses amenable to coarse demographic conditioning, Behavior tests demand context-sensitive reasoning about situational cues, and Coherence tests' bimodality suggests models either preserve inter-construct relationships well or break them entirely.

\subsubsection{Characterizing High and Low Scores}

To understand the differences between high and low simulation quality, we inspected agent outputs across configurations for the same test (see \ref{subsec:qual_methods}).

Consider the \textit{Interface Formality Effect on Disclosure} test~\cite{Leslie2010casual_vs_professional_interface}, where agents had to choose whether to self-disclose information on a casual (vs.\ professional) interface.
In a high-scoring configuration (Gemini-2.5-Flash, Demographic, SQ = 76.14), agents explicitly connected interface cues to perceived stakes: on a professional survey, ``providing false answers is the only way to complete the survey without compromising my personal information''; on a casual quiz, ``This is a silly online quiz, clearly just for fun. There's no real consequence to answering honestly.''
This rationale matched the empirical mechanism reported in the original study, in which study participants felt more comfortable disclosing information on a casual interface. In other words, disclosure behavior shifted with perceived consequences rather than reflecting fixed privacy preferences.

A low-scoring configuration for the same test (Gemini-2.5-Flash, Scenario-Primed Persona, SQ = 1.05) exhibited the reverse effect.
The agent treated the casual quiz as threatening (``I don't know who's asking these questions'') while trusting the ``professional'' survey interface.
The Scenario-Primed Persona strategy induced a global distrust policy that suppressed consideration of interface cues.

For Attitude tests, high-scoring configurations differentiated between related constructs (e.g., ``I worry about how much data is collected, but I still trust well-known sites''; Gemini-2.5-Flash, S\&P-Primed Persona, SQ = 79.62).
Agent rationales in low-scoring configurations collapsed into extreme, repetitive responses (e.g., ``I share anything online without really thinking about privacy'' repeated across items; Qwen3-Next-80B, Raw Persona, SQ = 24.64), reducing inter-agent variance and, thus, creating an unrealistically uniform population distribution.

For Coherence tests, the \textit{Privacy Structural Paths} test~\cite{malhotra2004_IUIPC} predicts that greater privacy concern should increase perceived risk and reduce willingness to disclose.
In a high-scoring configuration (GPT-5 with theory-informed prompting, SQ = 96.71), the agent explicitly linked constructs: ``because I am very concerned, I see more risk, so I am less willing to disclose even if I somewhat trust the site.''
A low-scoring configuration (GPT-5-Nano, SQ = 17.84) decoupled behavior from modeled constructs: ``I know I should care about security, but reusing simple passwords is easier.''

Overall, the SQ metric worked as intended: high-scoring simulations were context-sensitive and mechanism-consistent, while low-scoring simulations applied blunt global policies, produced repetitive responses, or reversed expected effect directions.

\subsection{RQ1: Do More Capable Models Produce Better Simulations?}
\label{sec:rq1}

\begin{figure}[t]
\centering
\includegraphics[width=\linewidth]{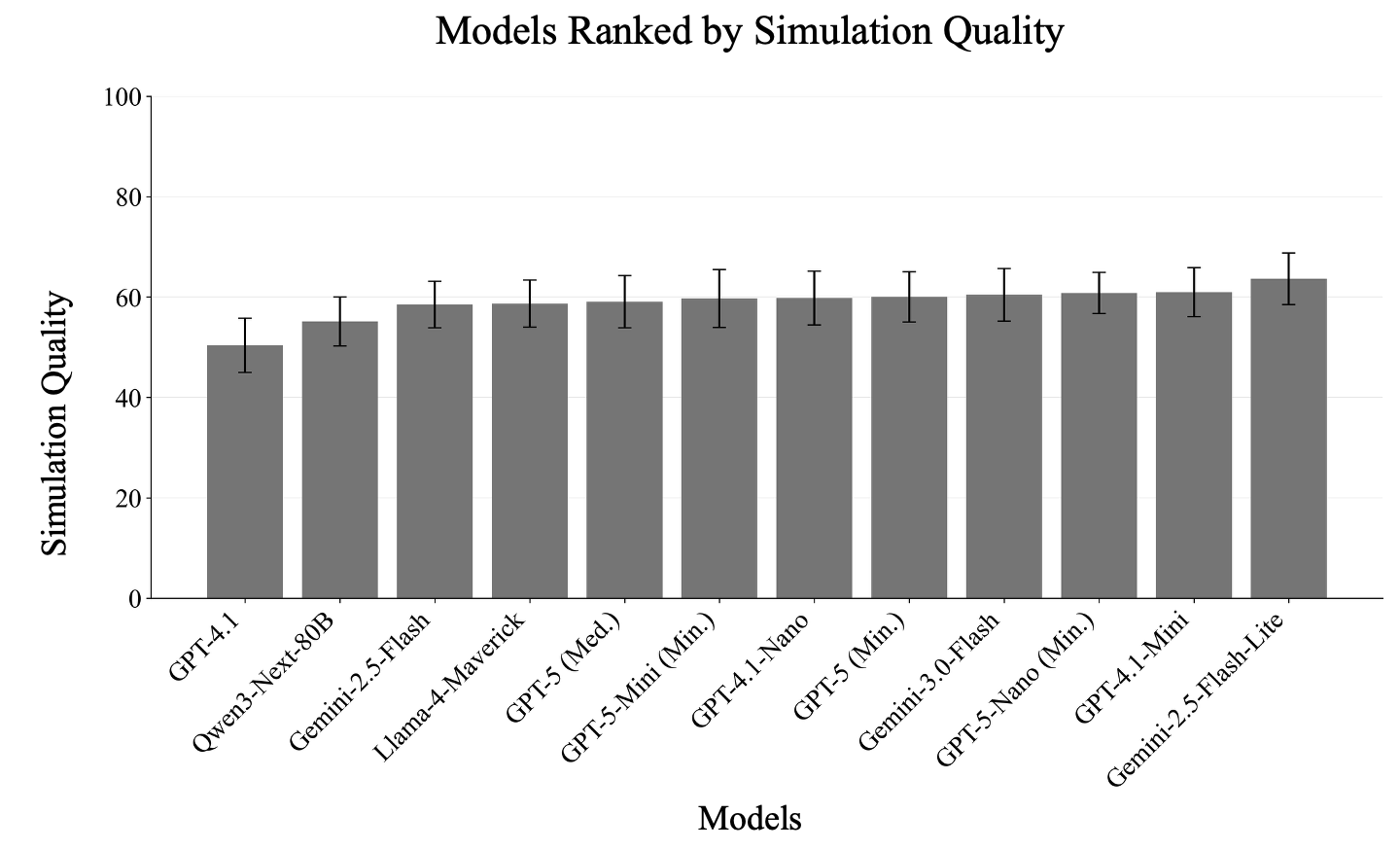}
\caption{Models ranked by average simulation quality across all tests. Error bars show standard error. The 13-point spread from GPT-4.1 (50.36) to Gemini-2.5-Flash-Lite (63.63) indicates model choice affects alignment, but no model achieves uniformly high fidelity.}
\label{fig:model-overall}
\end{figure}

\subsubsection{Model Differences Were Small Overall}

Figure~\ref{fig:model-overall} ranks all twelve models by average simulation quality across all 30 tests in \textsc{SP-ABCBench}.
The lowest-scoring model (GPT-4.1, 50.36) and highest-scoring model (Gemini-2.5-Flash-Lite, 63.63) differed by only 13 SQ points.
Model selection did not strongly affect average SQ, though specific models yielded higher scores on specific dimensions.

\begin{figure*}[t]
\centering
\includegraphics[width=0.8\textwidth]{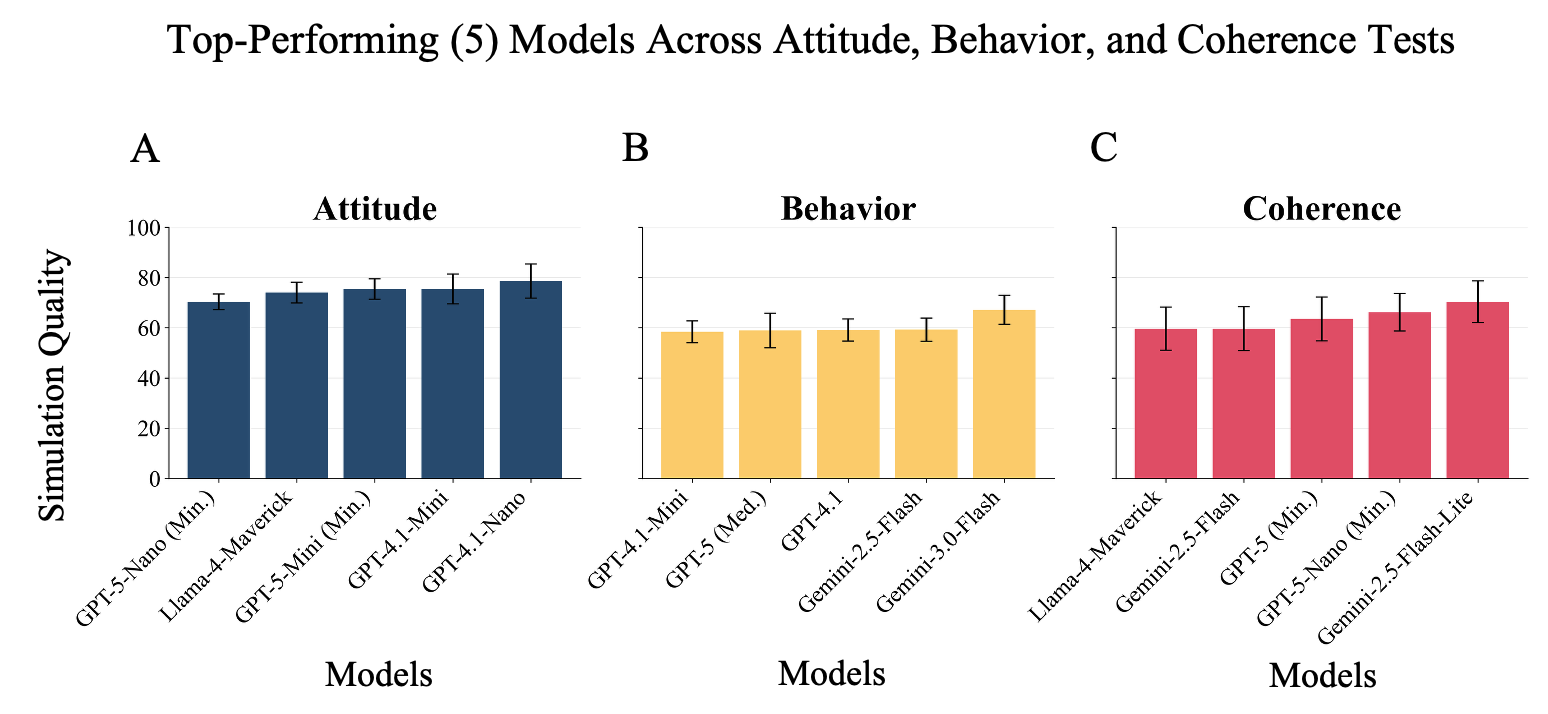}
\caption{Top-performing models across Attitude, Behavior, and Coherence tests. Rankings differ by dimension: GPT-4.1-Nano leads Attitude, Gemini-3.0-Flash leads Behavior, and Gemini-2.5-Flash-Lite leads Coherence. No single model dominates across all three dimensions.}
\label{fig:model-dimension}
\end{figure*}

Figure~\ref{fig:model-dimension} reports the top five models, based on average overall performance, within each dimension.
GPT-4.1-Nano led Attitude tests (mean SQ = 78.56); Gemini-2.5-Flash-Lite led Coherence tests (mean SQ = 70.32); Gemini-3.0-Flash led Behavior tests (mean SQ = 67.07), nearly 8 points ahead of the next-best model.
No single model led all three dimensions.

\subsubsection{Bigger, Newer, and Smarter Models Were Often Worse}

\begin{figure}[t]
\centering
\includegraphics[width=\linewidth]{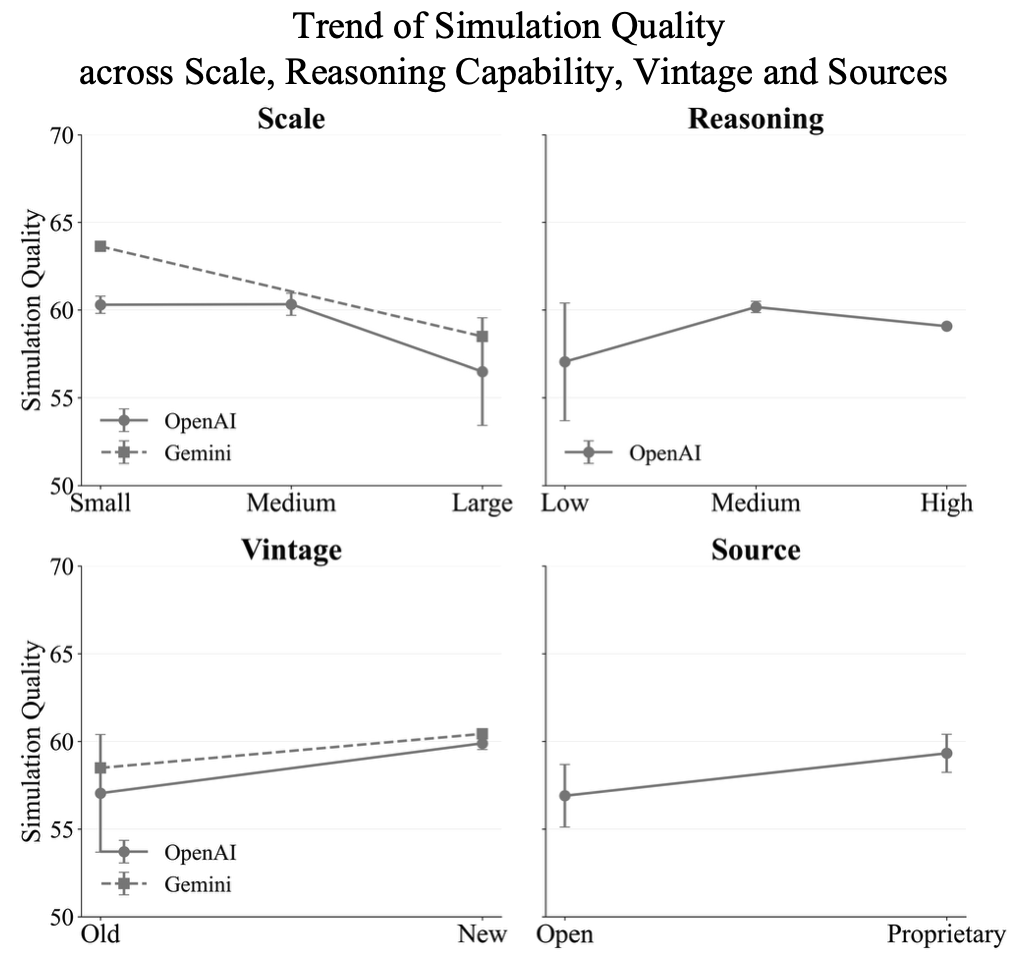}
\caption{Simulation quality trends across model characteristics. Scale does not monotonically improve simulation quality; medium reasoning performs comparably to minimal reasoning; newer generations yield small average gains; open-source models trail proprietary models only slightly.}
\label{fig:model-trends}
\end{figure}

Figure~\ref{fig:model-trends} groups models by scale, generation, reasoning capability, and source.

\textbf{Smaller models often outperformed larger counterparts.}
Within the OpenAI family, smaller variants (GPT-4.1-Nano, GPT-5-Nano) averaged 60.30 SQ, while larger variants (GPT-4.1, GPT-5, GPT-5 Medium) averaged 56.48 SQ---nearly 4 points lower.
Within-family comparisons showed the same pattern: GPT-4.1-Mini (60.96) surpassed GPT-4.1 (50.36) by over 10 points; Gemini-2.5-Flash-Lite (63.63) surpassed Gemini-2.5-Flash (58.49) by 5 points.
This is consistent with the ``hyper-accuracy distortion''~\cite{pmlr-v202-aher23a}, where greater capability may impair simulation of the heuristic-driven, suboptimal reasoning characteristic of real S\&P decisions (see Section~\ref{sec:discussion}).

\textbf{Newer models offered only modest SQ gains.}
GPT-5 variants averaged 59.89 SQ versus 57.04 for GPT-4.1 variants---under 3 points and frequently within standard error bounds.

\textbf{Extended reasoning did not help.}
GPT-5 (Medium Reasoning) scored 59.07 SQ, slightly \textit{worse} than GPT-5 (Minimal Reasoning) at 60.02.
Additional deliberation budget did not translate to improved population-level alignment.

\textbf{Open-source models matched proprietary models.}
Proprietary models averaged 59.33 SQ versus 56.90 for open-source models (Qwen3-Next-80B, Llama-4-Maverick). Nevertheless, Llama-4-Maverick (58.68) exceeded several proprietary alternatives, and the gap between open- and closed-source models was small relative to within-group variance.

\subsection{RQ2: Which Persona Construction Strategies Work Best?}
\label{sec:rq2}

We compared the four persona construction strategies from Section~\ref{subsubsec:PCS}: Demographic, Raw Persona, S\&P-Primed Persona, and Scenario-Primed Persona.
Persona strategy effects were dimension-dependent: e.g., strategies that improved SQ for Behavior tests often reduced SQ for Attitude and/or Coherence tests.

\subsubsection{No Single Strategy Performed Best Across All Dimensions}

\begin{figure}[t]
\centering
\includegraphics[width=\linewidth]{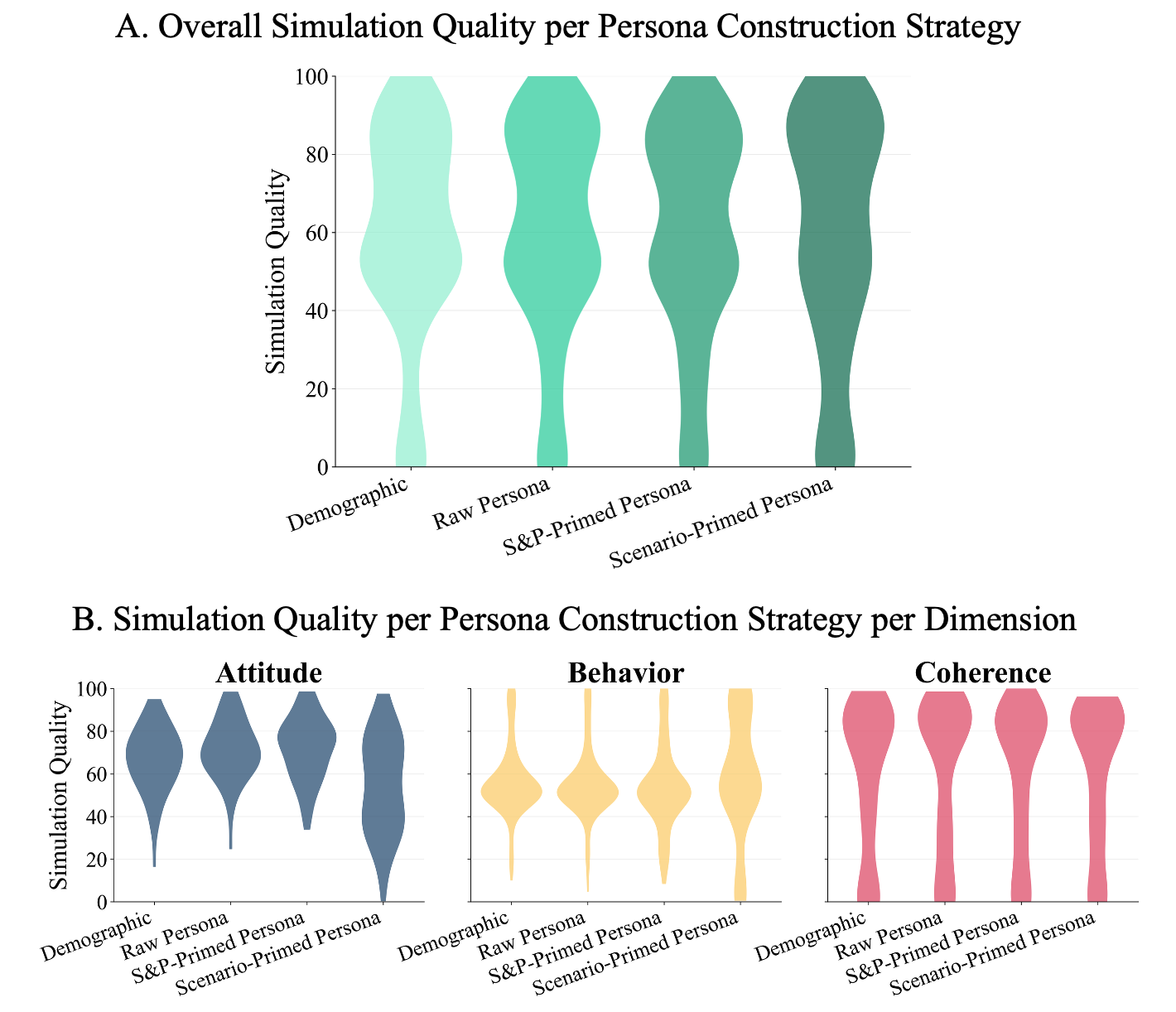}
\caption{Simulation quality by persona construction strategy. (A) Overall distributions have similar means but different shapes. (B) Dimension-specific patterns reveal trade-offs: Scenario-Primed Persona exhibits high variance on Attitude (including very low scores) but performs well on Behavior.}
\label{fig:pcs-violin}
\end{figure}

Figure~\ref{fig:pcs-violin} compares score distributions by persona strategy.
Overall means were similar (58.66--59.50), but dimension-specific patterns differed substantially.

\textbf{Attitude tests: S\&P-Primed best, Scenario-Primed worst.}
S\&P-Primed personas achieved the highest SQ for Attitude tests (71.87), with Raw Personas close behind (70.91).
Scenario-Primed Personas scored lowest (54.46)---a 17-point gap.
Providing too much scenario context for Attitude tests overly primed agent responses and reduced the population-level variance needed to match real world distributions.

Consider the OPC item ``Do you shred/burn your personal documents when you are disposing of them?''~\cite{buchanan2007_OPC1}.
S\&P-Primed Personas exhibited realistic variance across the five-point scale (counts: 2: 21, 3: 43, 4: 234, 5: 461), with 125 of 759 mentioning practical constraints like ``I don't have a shredder.''
Scenario-Primed Personas responded more unilaterally (counts: 3: 1, 4: 4, 5: 754), with only 1 of 759 mentioning lack of equipment.
The explicit emphasis on privacy-protective behaviors anchored agents to idealized responses.

\textbf{Behavior tests: Scenario-Primed best.}
Conversely, Scenario-Primed Personas yielded the highest SQ for Behavior tests (58.55), surpassing the next best Demographic personas (56.65) by 1.9 points.
Priming agents with scenario context helped them attend to specific experimental manipulations---interface formality, normative cues, institutional signals---that drove behavioral effects.

Consider the Policy Link test~\cite{Hoofnagle2014disclosure_rate_with_vs_without_privacy_policy}, which measures whether users disclose more when a privacy policy is visibly linked.
With S\&P-Primed Persona, agents provided fake information regardless of whether a policy was present (99--100 of 100), failing to reproduce the published effect.
With Scenario-Primed Persona, agents discriminated between conditions: 100 of 100 provided fake information without a policy link, but only 21 of 100 did so when the link was present---closely matching the published effect.
Scenario-Primed agents explicitly referenced the policy link as a trust signal (``I see a clear, prominently linked Privacy Policy... I'm comfortable using my real name''), whereas S\&P-Primed agents applied a blanket distrust heuristic.

\textbf{Coherence tests: Unaffected by persona strategy.}
Across persona construction strategies, Coherence test means ranged from 58.61 (Demographic) to 59.86 (Scenario-Primed)---a spread of only 1.25 points.
By contrast, Attitude tests spanned 17.41 points and Behavior tests spanned 4.56 points across the best- and worst strategies.
Because Coherence tests evaluate structural relationships (correlations, SEM paths, demographic differences) rather than absolute response levels, 
they may be less sensitive to global shifts in response tendency.

\subsubsection{Model $\times$ Persona Strategy Interactions}

\begin{figure}[t]
\centering
\includegraphics[width=\linewidth]{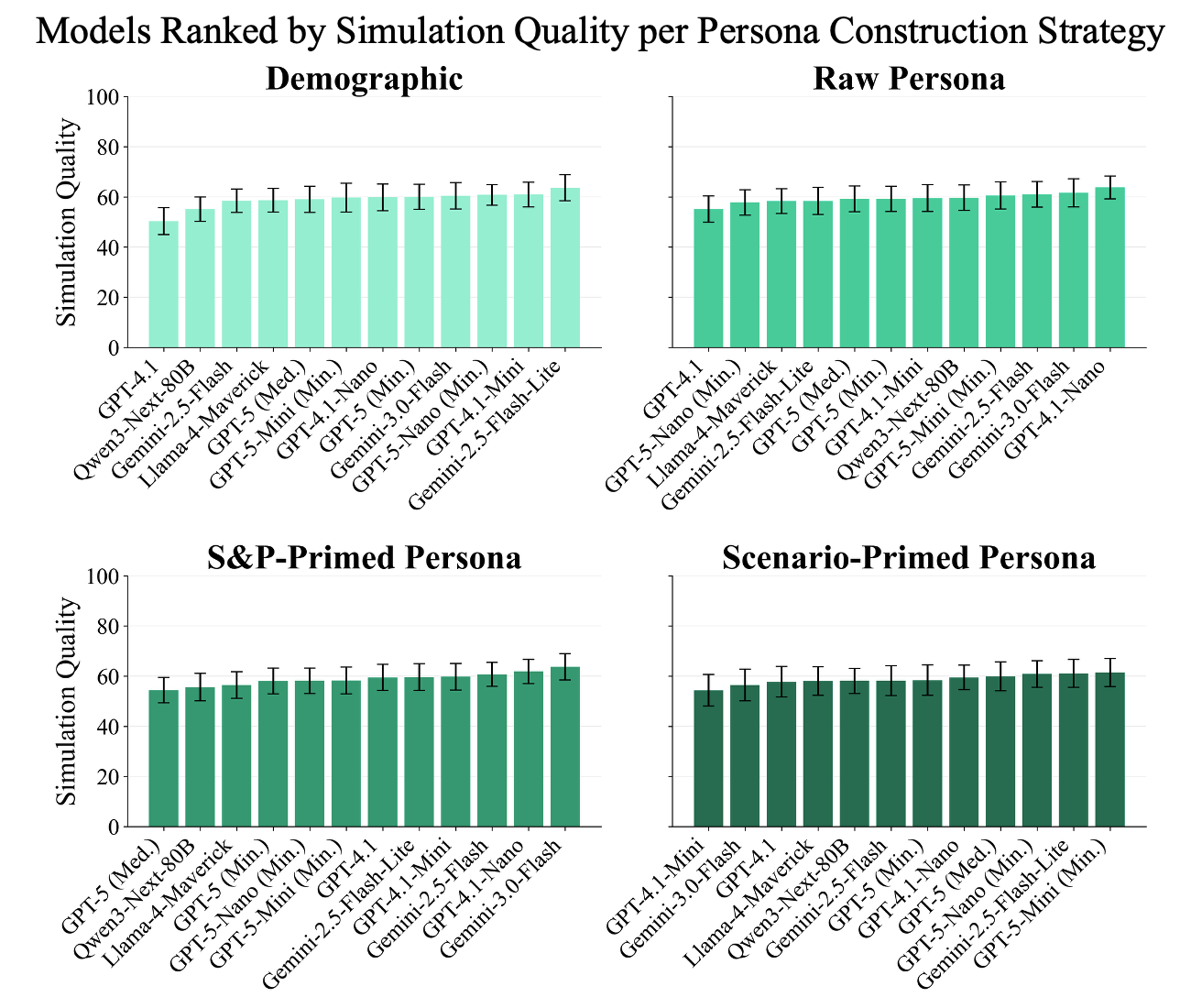}
\caption{Models ranked by simulation quality vary across persona strategies: GPT-4.1-Nano leads under Raw Persona; GPT-5-Mini leads under Scenario-Primed. Optimal persona choice depends on the model.}
\label{fig:pcs-models}
\end{figure}

Figure~\ref{fig:pcs-models} shows that model rankings shifted across persona strategies, indicating interaction effects.
GPT-5-Mini benefited from Scenario-Primed Persona, achieving the highest average SQ (61.41) among all model-persona configurations.
GPT-4.1-Mini showed the opposite pattern (Demographic: 60.96 vs.\ Scenario-Primed: 54.34).
This interaction effect was not related to model vintage: GPT-4.1 (base) benefited substantially from Scenario-Primed personas (+7.4 points), while GPT-4.1-Mini was harmed by it ($-$6.6 points).

For Attitude tests, Scenario-Primed personas were particularly ineffective for larger models.
GPT-5 (Minimal Reasoning) with Scenario-Primed Persona dropped from 67.55 average SQ (Demographic) to 40.35---a 27-point reduction.
Similar fall-offs occurred for GPT-4.1-Mini ($-$28.8), Qwen3-Next-80B ($-$24.0), and GPT-5 (Medium Reasoning) ($-$18.1).
GPT-4.1-Nano --- a smaller model --- bucked this trend, actually improving with Scenario-Primed personas (78.56 to 81.88), suggesting smaller models may have been less susceptible to over-priming.

These interactions imply that persona construction must be tailored to both the model and the simulation objective.
We discuss this in Section~\ref{subsubsec:pcs_trade_off}.

\subsection{RQ3: How Did Prompting Approach Impact Simulation Quality?}
\label{sec:rq3}

We compared baseline and theory-informed prompting (Section~\ref{subsec:theory_informed_prompting_experiment}) using four widely-used models that span our overall ranking: GPT-4.1 (rank 12), Gemini-2.5-Flash (rank 10), GPT-5 (Minimal Reasoning, rank 5), and Gemini-2.5-Flash-Lite (rank 1).

\subsubsection{Overall Effect}

\begin{figure}[t]
\centering
\includegraphics[width=\linewidth]{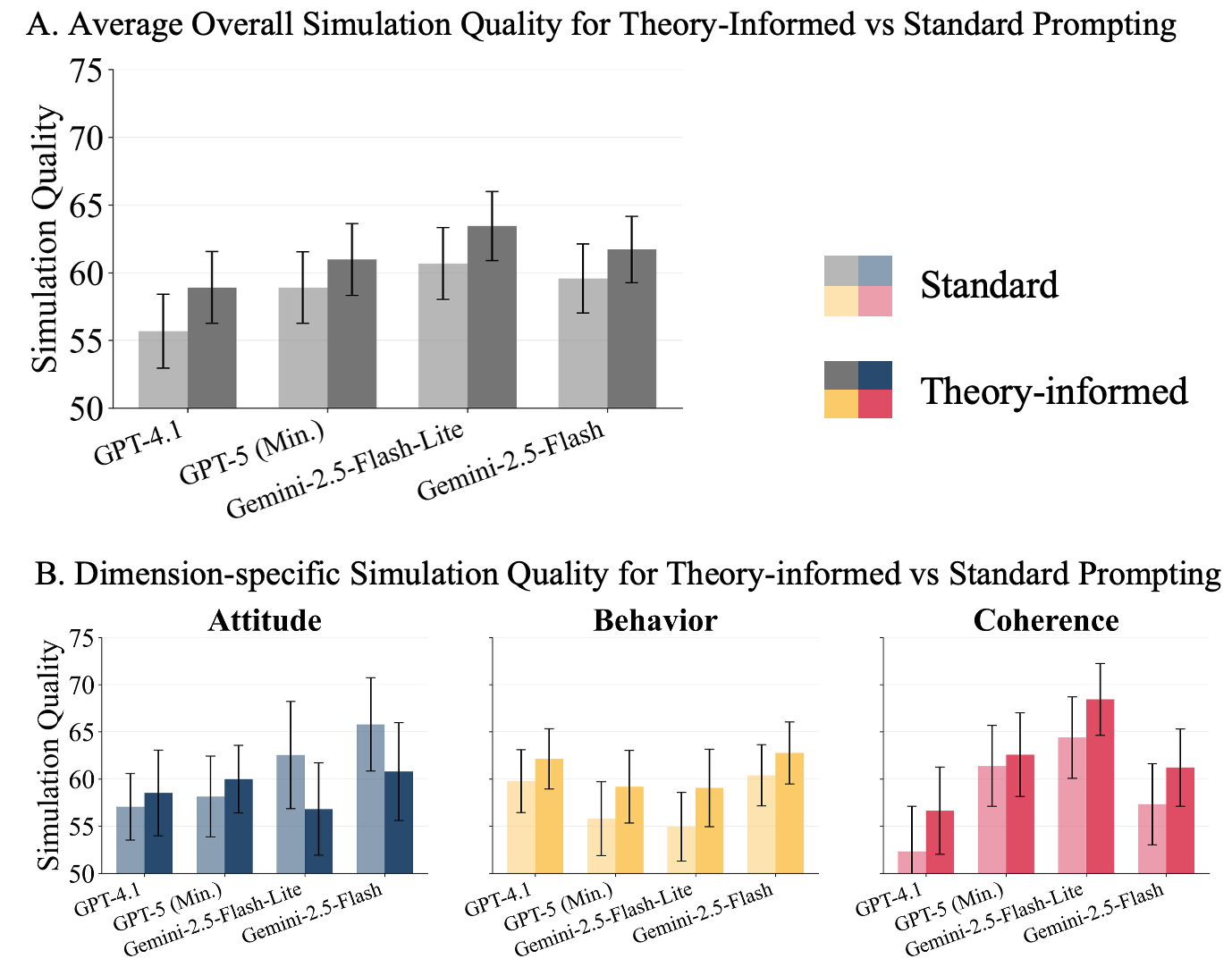}
\caption{Effect of theory-informed prompting on simulation quality across four models. (A) Overall improvement of 2.55 points averaged across models. (B) Dimension-specific effects: Behavior (+3.05) and Coherence (+3.36) improve consistently, while Attitude slightly degrades ($-$1.85) on average.}
\label{fig:tip-effect}
\end{figure}

Figure~\ref{fig:tip-effect} shows that theory-informed prompting improved average SQ across all four models (GPT-4.1: +3.23, GPT-5: +2.07, Gemini-2.5-Flash-Lite: +2.76, Gemini-2.5-Flash: +2.14), with an overall gain of 2.55 points.
However, gains were dimension-dependent.
Behavior (+3.05) and Coherence (+3.36) improved consistently across models, while Attitude slightly degraded on average ($-$1.85).
This Attitude degradation was model-dependent: Gemini models showed substantial drops (Gemini-2.5-Flash-Lite: $-$5.71, Gemini-2.5-Flash: $-$4.99), while OpenAI models showed slight gains (GPT-4.1: +1.47, GPT-5: +1.83).

\subsubsection{The Mechanism: Trade-off Reasoning}

Theory-informed prompting shifted agent reasoning from simple global policies to context-specific cost-benefit trade-offs.
Under standard prompting, agents applied blanket rules: e.g., ``I value honesty, but I also guard my privacy\ldots the privacy-preserving move is to skip.''
Under theory-informed prompting, agents instead weighed situational costs against benefits: e.g., ``It's a sensitive, potentially self-incriminating question with no clear \textit{benefit} to me personally.''
This shift was quantifiable: on the Normative Information test~\cite{Acquisti2012disclosure_rate_high_vs_missing_norm}, the term ``benefit'' appeared in 13 of 100 agent rationales under standard prompting versus 61 of 100 under theory-informed prompting---a 4.7$\times$ increase.
Both groups of agents arrived at similar decisions (79/100 vs.\ 90/100 chose not to disclose), but theory-informed agents grounded their choices in explicit cost-benefit calculus rather than general privacy preferences.

This trade-off reasoning---encouraged by Privacy Calculus and Bounded Rationality theory~\cite{dinev2006extended, simon1955behavioral}---explains why Behavior and Coherence tests improved: real users weigh situational costs and benefits when \textit{acting}, and theory-informed agents better reproduced this \textit{deliberation}.

However, the same mechanism degraded Attitude tests, where real respondents tend toward more idealized positions.
Consider the SA-6 attitude item ``I am extremely motivated to take all the steps needed to keep my online data and accounts safe''~\cite{faklaris2019_SA6}.
Under standard prompting, agents responded categorically: ``My career in tech\ldots makes data and account security a top priority. I'm extremely motivated'' (score: 5/5).
Under theory-informed prompting, the same persona hedged: ``I balance security with convenience and practicality, so I'd prioritize the most impactful steps rather than every single one'' (score: 3/5).
This pattern was systematic: in Gemini-2.5-Flash, the trade-off marker ``convenience'' appeared in 75 of 209 responses under theory-informed prompting versus only 2 of 209 under standard prompting.
The resulting less-extreme scores reduced alignment with real attitude distributions, which tend to skew toward the idealistic end---a manifestation of the privacy paradox~\cite{barth2017privacy}.
In short, theory-informed prompting made agents reason more like people \textit{act}, not like people \textit{say they would act}.

\subsubsection{Persona Strategy $\times$ Prompting Interactions}

Table~\ref{tab:tip-pcs-full} reports the full breakdown by persona strategy and dimension, averaged across all four models.
All four persona strategies improved overall, with Demographic (+3.64) and Raw Persona (+3.58) gaining most, and S\&P-Primed gaining least (+0.95).

Dimension-specific trade-offs emerged.
For Behavior tests, three of four strategies improved, with Raw Persona showing the largest gain (+5.22); Scenario-Primed Personas, which already had strong baseline Behavior performance, showed negligible change ($-$0.08).
For Attitude tests, the pattern reversed: strategies with stronger baselines (Raw Persona, S\&P-Primed) degraded ($-$6.81, $-$2.12), while the weakest baseline (Scenario-Primed) improved substantially (+6.31).
For Coherence tests, Demographic and Raw Persona improved most (+6.63, +5.15), while S\&P-Primed slightly degraded ($-$0.78).

A consistent pattern emerged across dimensions: theory-informed prompting helped most where baseline performance was weakest, and occasionally degraded configurations that already performed well.
Figure~\ref{fig:tip-pcs} in Appendix~\ref{app:additional_figure} provides a visualized comparison.

\begin{table}[t]
\centering
\caption{Theory-informed prompting effects by persona strategy and dimension, averaged across four models (GPT-4.1, GPT-5, Gemini-2.5-Flash-Lite, Gemini-2.5-Flash). Negative values indicate degradation.}
\label{tab:tip-pcs-full}
\small
\begin{tabular}{llrrr}
\toprule
\textbf{Dim.} & \textbf{Strategy} & \textbf{Std.} & \textbf{Theory} & \textbf{$\Delta$} \\
\midrule
\multirow{4}{*}{Overall}
& Demographic & 58.12 & 61.76 & +3.64 \\
& Raw Persona & 58.41 & 61.99 & +3.58 \\
& S\&P-Primed & 59.44 & 60.40 & +0.95 \\
& Scenario-Primed & 58.85 & 60.88 & +2.02 \\
\midrule
\multirow{4}{*}{Att.}
& Demographic & 57.78 & 53.00 & $-$4.78 \\
& Raw Persona & 65.77 & 58.96 & $-$6.81 \\
& S\&P-Primed & 69.70 & 67.59 & $-$2.12 \\
& Scenario-Primed & 50.25 & 56.56 & +6.31 \\
\midrule
\multirow{4}{*}{Beh.}
& Demographic & 56.45 & 59.07 & +2.63 \\
& Raw Persona & 56.26 & 61.47 & +5.22 \\
& S\&P-Primed & 56.45 & 60.89 & +4.44 \\
& Scenario-Primed & 61.75 & 61.67 & $-$0.08 \\
\midrule
\multirow{4}{*}{Coh.}
& Demographic & 59.45 & 66.07 & +6.63 \\
& Raw Persona & 58.03 & 63.19 & +5.15 \\
& S\&P-Primed & 58.90 & 58.12 & $-$0.78 \\
& Scenario-Primed & 59.03 & 61.45 & +2.42 \\
\bottomrule
\end{tabular}
\end{table}

\section{Discussion}
\label{sec:discussion}

Our evaluation reveals both the promise and limitations of using LLM agent simulations to reproduce population-level S\&P patterns.
Some configurations achieved high alignment---scoring above 95 SQ out of 100 on individual tests---while average performance remained moderate (50--64 SQ across configurations).
Simulation quality depends critically on methodological choices, and substantial room for improvement exists.

\subsection{What Our Results Reveal}

\subsubsection{Current Capabilities}

The best-performing configuration achieved 63.63 average simulation quality across all tests on \textsc{SP-ABCBench}, but this average performance masks considerable variability: GPT-5 with theory-informed prompting scored 96.71 on the Privacy Structural Paths test, while other configurations fell below 20 SQ.
An SQ score around 60---representative of many configurations---typically means simulations capture effect directions but miss magnitudes or compress population variance.
For instance, on the OPC attitude test, configurations scoring around 60 reproduced subscale rank-ordering but responded too uniformly, missing the heterogeneity typically present in human populations.

These patterns suggest current LLM agent simulations may achieve what prior work on generative agent simulations calls ``believability''---i.e., plausible outputs on inspection---without necessarily achieving the ``accuracy'' required for quantitative prediction~\cite{park2024generative}.
For applications requiring only plausibility, current methods may suffice; for accurate effect sizes, significant improvements are needed.

\subsubsection{Model Scale, Vintage, and Origin Does Not Predict Quality}

Counterintuitively, larger and newer models do not consistently outperform smaller ones: GPT-4.1-Nano outperforms GPT-4.1; Gemini-2.5-Flash-Lite outperforms Gemini-2.5-Flash; extended reasoning in GPT-5 offers no improvement.
This aligns with what prior work calls the ``hyper-accuracy distortion''~\cite{pmlr-v202-aher23a}, where increased model capability may impair simulation of human behaviors that deviate from normative rationality.
S\&P decisions are secondary to primary goals~\cite{whitten1999johnny, adams1999users}, preventive rather than rewarding~\cite{anderson2001information, campbell2003economic}, and oriented toward abstract threats~\cite{friedman2002informed, wash2010folk}---characteristics that larger models optimized for consistency may handle poorly by overproducing idealized or overly rational responses and compressing behavioral variance.

\subsubsection{Persona Construction Strategy Trade-offs}
\label{subsubsec:pcs_trade_off}

Persona construction strategies show dimension-specific effects.
For example, Scenario-Primed Persona improves Behavior tests (+1.9 points) but degrades Attitude tests by 17 points relative to the best strategy.
On the Policy Link test, Scenario-Primed agents discriminated between conditions (21\% vs.\ 100\% fake information with/without policy link), matching the published effect; S\&P-Primed agents applied blanket distrust regardless of condition.
Conversely, on OPC items (an Attitude test), Scenario-Primed Personas output unilateral responses with little inter-agent variety, while S\&P-Primed Persona preserved realistic variance.
Persona construction must be tailored to simulation objectives.

\subsubsection{Theory-Informed Prompting}

Theory-informed prompting grounded in Privacy Calculus and Bounded Rationality~\cite{culnan1999information, dinev2006extended, simon1955behavioral} improved behavioral simulation by 3.05 points on average, with larger gains for weaker baselines.
Agents shifted from global decision heuristics (``I don't share personal information online'') to context-specific trade-offs (``I'll share some things but not others depending on who might see'').
However, it degraded Attitude tests on average ($-$1.85), likely because trade-off reasoning reduced the idealistic response patterns that characterize real attitude data (Section~\ref{sec:rq3}).
Coherence tests improved comparably to Behavior (+3.36), suggesting that trade-off reasoning also strengthened structural relationships between constructs.

The Privacy Calculus and Bounded Rationality framing we tested are two of many possible approaches.
Other frameworks---Contextual Integrity~\cite{nissenbaum2004privacy}, SPAF~\cite{das2022security}---may prove effective for different objectives.
\textsc{SP-ABCBench} enables systematic evaluation of such alternatives.

\subsection{Implications for Practice}

Prior work has proposed using LLM simulations as ``crash dummies'' for S\&P testing~\cite{asaditoward}, for red-teaming~\cite{asadi2025personas}, and for safety simulation~\cite{zhouhaicosystem}.
Our findings provide empirical grounding for these applications.

\subsubsection{When LLM Simulations May (Not) Be Effective}

\textbf{Hypothesis generation.}
Using simulations to brainstorm user concerns or identify candidate failure modes only requires simulations to be ``believable'' \cite{park2024generative}.
Current methods may be believable enough to effectively complement expert intuition, though outputs should remain suggestive --- not predictive.

\textbf{Adversarial testing.}
Using agents as proxies for safety testing assumes realistic S\&P attitudes and behaviors.
Our findings raise caution: agents often over-applied security heuristics or exhibited idealized behavior.
On the Control Paradox test, where real users paradoxically disclosed \textit{more} under explicit control, some configurations reversed the effect entirely.

\textbf{Quantitative estimation.}
Using simulations to estimate adoption rates or effect sizes for design decisions requires accuracy that current methods do not reliably achieve.

\subsubsection{Recommendations}

\textbf{Match persona construction strategies to goals.}
Use Scenario-Primed for behavioral simulations; use less constrained personas for attitudinal distributions.

\textbf{Validate configuration parameters on \textsc{SP-ABCBench}.}
Do not assume larger models perform better.
Benchmark on task-relevant tests before deployment.


\textbf{Integrate with expert workflows.}
Simulations may work best in ways that expand the space of considered scenarios in threat modeling~\cite{shostack2014threat, kohnfelder1999threats} or PIAs~\cite{clarke2009privacy, bogucka2024ai}, with experts judging plausibility of outputs and prioritizing risks.

\subsection{Limitations and Future Work}

\textsc{SP-ABCBench} covers primarily Western, U.S.-based studies and emphasizes static, one-step scenario-based outcomes. Expanding to cross-cultural studies and multi-step tasks is essential for generalizability.
Our evaluation used prompt-based simulation without fine-tuning or memory.
Promising approaches we did not test include interview-based persona construction (shown to replicate users with 85\% accuracy in other domains~\cite{park2024generative}) and fine-tuning on behavioral data~\cite{kolluri2025finetuning}.
More generally, closing the gap between current performance and reliable simulation will require advances beyond model scaling.
\textsc{SP-ABCBench} provides a foundation for evaluating such advances, and we invite the community to benchmark new methods as they emerge.
\section{Conclusion}
\label{sec:conclusion}

Motivated by recent calls to use LLMs as proxies for people in red teaming and safety testing \cite{park2022social,zhouhaicosystem,zhou2025safeagent,asaditoward}, we evaluated how well current LLMs reproduce population-level security and privacy (S\&P) attitudes and behaviors.
We introduced \textsc{SP-ABCBench}, a benchmark of 30 tests derived from 15 validated studies of users' S\&P attitudes and behaviors.
Across twelve models, average simulation quality (SQ) ranged from 50 to 64 out of 100, leaving substantial room for improvement.
Larger or newer models did not consistently perform better, and sometimes performed worse. Still, certain configurations-specific combinations of model, persona construction, and prompting—achieved high SQ on individual tests (26.94\% scored 80+; 13.33\% scored 90+).
\textsc{SP-ABCBench} provides a calibrated foundation for identifying what works and tracking progress as simulation methods evolve.
We release the benchmark, evaluation code, and reasoning traces to support reproducibility and extension of \textsc{SP-ABCBench} by the S\&P research community.

\cleardoublepage
\appendix
\section*{Ethical Considerations}

This work evaluates large language models as simulators of population-level S\&P attitudes and behaviors using benchmarks derived from prior human-subject studies.  
We do not collect new data from human participants, and all empirical targets are drawn from previously published, peer-reviewed studies conducted under their original ethical approvals.  
Synthetic populations are constructed by sampling publicly available U.S. Census marginals, without representing or re-identifying real individuals.  
LLM agents are treated as abstract simulators rather than proxies for specific people, and our analysis explicitly focuses on aggregate patterns rather than individual prediction.  
A potential ethical risk is the misuse of simulations to replace or overrule human-subject research in high-stakes S\&P decisions.  
Our results show that current simulations exhibit only moderate alignment with empirical data, and we explicitly caution against deploying them for decision-critical or normative judgments.  
There is a risk that simulated outputs could reinforce stereotypes or overgeneralized assumptions about demographic groups if interpreted uncritically.  
To mitigate this risk, we restrict demographic attributes to coarse categories used in the source studies and evaluate alignment only at the population level.  
Another concern is that models trained on web data may encode historical biases about security awareness, privacy concern, or ``rational'' behavior.  
Our benchmark is designed to reveal such misalignments rather than to endorse simulated behavior as desirable or correct.  
We release \textsc{SP-ABCBench} as an evaluation and calibration tool, not as a system for predicting or prescribing user behavior.
All reported results include clear limitations and are framed to discourage overconfidence in LLM-based population simulation.  
We believe this work supports responsible use by clarifying where current methods fail, thereby reducing the risk of inappropriate reliance on simulations in security and privacy practice.  
\cleardoublepage

\section*{Open Science}

We commit to open science by releasing all artifacts necessary to evaluate, reproduce, and extend the contributions of this paper.  
All released materials will be provided via an anonymous GitHub repository~\footnote{\url{https://anonymous.4open.science/r/ABCBench-B808/README.md}}.

The repository will include the full \textsc{SP-ABCBench} benchmark, including formal definitions of all Attitude, Behavior, and Coherence tests and their mappings to source human-subject studies.  
We will release the complete evaluation pipeline code, covering demographic sampling, persona construction, agent prompting, response aggregation, and simulation quality scoring.  
All data generated in our experiments will be provided, including raw agent responses for survey and scenario simulations, aggregated results, computed metrics, and final scores.  
We will also release all model reasoning traces collected during evaluation to support transparency and further analysis.  

The repository will contain scripts and configuration files that reproduce every experiment reported in the paper, including model lists, persona strategies, and evaluation settings.  
Exact prompts and prompting templates used for persona generation and task inference are documented in Appendix~\ref{app:prompts}.  
All generation parameters (e.g., temperature and sampling settings) are also specified in Appendix~\ref{app:demographic_sampling} and Appendix~\ref{app:parameter_settings}.  

To support reproducibility, we will document software dependencies, library versions, and instructions for running the full pipeline end-to-end.  
Where external APIs are required (e.g., for LLM requests), we will provide clear interfaces and instructions for substituting credentials while keeping all other components identical.  

We do not release any private or sensitive human data, as all empirical targets are drawn from previously published studies and all simulation outputs are synthetic.  
Together, these artifacts enable reviewers and future researchers to fully audit our methodology, reproduce our results, and benchmark new models or simulation techniques against \textsc{SP-ABCBench}.  
\cleardoublepage

\bibliographystyle{plain}
\bibliography{reference}

\section*{Appendix}     
\section{Demographic Sampling}
\label{app:demographic_sampling}

We generate demographic profiles by sampling from a filtered ACS PUMS pool. For each instrument, we match the target marginal distributions reported in the corresponding source study. We use a fixed random seed of 42 for reproducibility. IUIPC, SeBIS, SA-6, and the Behavior tests use robust stratified sampling; OPC uses a hybrid procedure (sex-stratified sampling with synthetic age generation).

\subsection{Pseudocode}
\label{app:demographic_sampling:pseudocode}

\begin{algorithm}[H]
\caption{Robust stratified demographic sampling (IUIPC, SeBIS, SA-6, Behavior tests).}
\label{alg:demographic_stratified}
\small
\begin{algorithmic}[1]
\Require Source pool $D$ (filtered ACS PUMS), target size $N$, target marginals for sex/education/age, RNG seed $s{=}42$
\State Initialize RNG $\mathcal{R} \leftarrow \textsc{Seed}(s)$
\State Derive stratification keys for each record in $D$:
\Statex \hspace{1.2em} $k_{\textit{sex}} \leftarrow \textsc{MapSex}(\texttt{SEX})$; \;
$k_{\textit{edu}} \leftarrow \textsc{MapEdu}(\texttt{SCHL})$; \;
$k_{\textit{age}} \leftarrow \textsc{BinAge}(\texttt{AGEP})$
\State Enumerate all strata $\mathcal{S} = k_{\textit{sex}} \times k_{\textit{edu}} \times k_{\textit{age}}$
\State Compute target stratum proportions $p(s) = p_{\textit{sex}} \cdot p_{\textit{edu}} \cdot p_{\textit{age}}$ for each $s \in \mathcal{S}$
\State Compute ideal allocations $\hat{n}(s) \leftarrow N \cdot p(s)$
\State Set base allocations $n(s) \leftarrow \lfloor \hat{n}(s) \rfloor$ and remainders $r(s) \leftarrow \hat{n}(s)-n(s)$
\State Distribute remaining slots $\Delta \leftarrow N - \sum_{s} n(s)$ to strata with largest $r(s)$ (Hamilton method)
\State Initialize result $R \leftarrow \emptyset$, used indices $U \leftarrow \emptyset$
\ForAll{$s \in \mathcal{S}$ with $n(s)>0$}
    \State Let $D_s \leftarrow \{x \in D : \textsc{Key}(x)=s\}$
    \State Sample $m \leftarrow \min(n(s), |D_s|)$ records from $D_s$ without replacement using $\mathcal{R}$
    \State $R \leftarrow R \cup \textsc{Sample}(D_s, m)$; update $U$ with sampled indices
\EndFor
\If{$|R| < N$} \Comment{Top-up for strata shortfalls}
    \State $k \leftarrow N-|R|$
    \State $D_{\textit{rem}} \leftarrow D \setminus U$
    \If{$|D_{\textit{rem}}| \ge k$}
        \State $R \leftarrow R \cup \textsc{Sample}(D_{\textit{rem}}, k)$ without replacement
    \Else
        \State $R \leftarrow R \cup \textsc{Sample}(D, k)$ with replacement
    \EndIf
\EndIf
\State \textbf{return} $R$
\end{algorithmic}
\end{algorithm}

\begin{algorithm}[H]
\caption{Hybrid demographic sampling (OPC).}
\label{alg:demographic_hybrid}
\small
\begin{algorithmic}[1]
\Require Source pool $D$ (filtered ACS PUMS), target size $N$, target sex proportions, target age $(\mu,\sigma,[a_{\min},a_{\max}])$, RNG seed $s{=}42$
\State Initialize RNG $\mathcal{R} \leftarrow \textsc{Seed}(s)$
\State Initialize result $R \leftarrow \emptyset$
\ForAll{sex categories $g$}
    \State $n_g \leftarrow \textsc{Round}(N \cdot p_{\textit{sex}}(g))$
    \State $D_g \leftarrow \{x \in D : \textsc{MapSex}(\texttt{SEX}(x))=g\}$
    \State $R \leftarrow R \cup \textsc{Sample}(D_g, n_g)$ with replacement using $\mathcal{R}$
\EndFor
\State If $|R| > N$, downsample to size $N$ using $\mathcal{R}$
\State Generate ages $a \sim \mathcal{N}(\mu,\sigma^2)$ for each record in $R$
\State Clip ages to $[a_{\min},a_{\max}]$ and cast to integers; store as \texttt{generated\_age}
\State \textbf{return} $R$
\end{algorithmic}
\end{algorithm}

\subsection{Post-processing}
\label{app:demographic_sampling:postprocessing}

We export each sampled record as a demographic profile containing \texttt{sex}, \texttt{age} (or \texttt{generated\_age} for OPC), \texttt{education} (mapped from \texttt{SCHL}), and \texttt{income} (from \texttt{PINCP}). For IUIPC, we additionally assign \texttt{scenario\_type} by shuffling an equal split of labels (A/B) under the same seed.

\section{Simulation Settings}
\label{app:prompts}

This appendix reports (i) the prompts used for persona construction, (ii) the prompts for survey/scenario simulation, and (iii) the hyper-parameters for API calls. We show prompts verbatim with placeholders in \texttt{monospace}.

\subsection{Persona Generation Prompts}
\label{app:persona_prompts}

\paragraph{P1: Demographic (no generated description).}
Agents receive only demographic attributes; no LLM call is made.
\begin{quote}
\small\ttfamily
Demographic Data:\\
sex: \texttt{<SEX>}\\
age: \texttt{<AGE>}\\
education: \texttt{<EDUCATION>}\\
income: \texttt{<INCOME>}
\end{quote}

\paragraph{P2: Raw Persona (biographical augmentation).}
We generate a 120--150 word persona description from demographics.
\begin{quote}
\small\ttfamily
\textbf{System:}\\
You are a creative writer whose job is to turn dry demographic data into a detailed, reasonable, and realistic user persona.\\
Based on the given demographic profile, write a 120-150 words description of a single person, include relevant details, such as name, race, personality, hobbies, occupation, most memorable experience, current life goal, family situation, etc.\\
The description should be second person, start with something like `Your name is ...'.\\
Ensure the description matches the demographic data, and is reasonable so that a reader feels meeting a genuine person.\\
\\
\textbf{User:}\\
Demographic Data:\\
sex: \texttt{<SEX>}\\
age: \texttt{<AGE>}\\
education: \texttt{<EDUCATION>}\\
income: \texttt{<INCOME>}\\
Please think step-by-step.
\end{quote}

\paragraph{P3: S\&P-Primed Persona (domain-oriented augmentation).}
We generate a 120--150 word persona description and explicitly ask for security and privacy attitudes, behaviors, and decisions.
\begin{quote}
\small\ttfamily
\textbf{System:}\\
You are a creative writer whose job is to turn dry demographic data into a detailed, reasonable, and realistic user persona.\\
Based on the given demographic profile, write a 120-150 words description of a single person, include relevant details, such as name, race, personality, hobbies, occupation, most memorable experience, current life goal, family situation, etc.\\
Specifically, describe this person's security and privacy attitudes, behaviors, and decisions.\\
The description should be second person, start with something like `Your name is ...'.\\
Ensure the description matches the demographic data, and is reasonable so that a reader feels meeting a genuine person.\\
\\
\textbf{User:}\\
Demographic Data:\\
sex: \texttt{<SEX>}\\
age: \texttt{<AGE>}\\
education: \texttt{<EDUCATION>}\\
income: \texttt{<INCOME>}\\
Please think step-by-step.
\end{quote}

\paragraph{P4: Scenario-Primed Persona (task-conditioned augmentation).}
We generate a 120--150 word persona description conditioned on the target test (or a specific behavioral scenario).
\begin{quote}
\small\ttfamily
\textbf{System:}\\
You are a creative writer whose job is to turn dry demographic data into a detailed, reasonable, and realistic user persona.\\
Based on the given demographic profile, write a 120-150 words description of a single person, include relevant details, such as name, race, personality, hobbies, occupation, most memorable experience, current life goal, family situation, etc.\\
\texttt{<TASK\_SPECIFIC\_INFO>}\\
The description should be second person, start with something like `Your name is ...'.\\
Ensure the description matches the demographic data, and is reasonable so that a reader feels meeting a genuine person.\\
\\
\textbf{User:}\\
Demographic Data:\\
sex: \texttt{<SEX>}\\
age: \texttt{<AGE>}\\
education: \texttt{<EDUCATION>}\\
income: \texttt{<INCOME>}\\
Please think step-by-step.
\end{quote}

\paragraph{Task-specific info blocks.}
For attitude and coherence tests, we condition personas on construct-relevant traits (e.g., IUIPC, SeBIS, SA-6, OPC). For behavioral tests, we condition on the specific scenario template indexed by \texttt{<TASK\_ID>}. \texttt{<TASK\_SPECIFIC\_INFO>} is instantiated by one of the following blocks:

\begin{quote}
\small\ttfamily
\textbf{IUIPC:} Sense of Control Over Personal Information; Awareness of Privacy Practices; Concern About Data Collection; Trust in Online Companies; Perception of Privacy Risk; Information Disclosure Intention.\\
\\
\textbf{SeBIS:} Physical Device Security; Password Management Practices; Proactive Online Awareness; Software and System Maintenance.\\
\\
\textbf{SA-6:} Proactive Interest in Security; Security Motivation and Diligence; Self-Perceived Security Knowledge.\\
\\
\textbf{OPC:} General Caution and Protective Habits; Technical Privacy Management; Overall Level of Privacy Concern; Apprehension about Digital Communications; Trust in Online Entities.\\
\\
\textbf{Behavior Scenarios:} Scenario-conditioned behavioral tendencies for \texttt{<TASK\_ID>} (one of 10 scenario families).
\end{quote}

\subsection{Simulation Prompts}
\label{app:simulation_prompts}

\paragraph{Simulation system prompt.}
All survey and scenario simulations use the following system instruction:
\begin{quote}
\small\ttfamily
You are a person of the given persona, and you need to complete a survey (or make a decision in the scenario provided). Respond authentically as this person would in real life, not as an AI.
\end{quote}

\paragraph{Survey simulation user prompt.}
\begin{quote}
\small\ttfamily
\texttt{<PERSONA\_PROMPT>}\\
\\
\texttt{<SURVEY\_PROMPT>}\\
For each question, think step-by-step and choose one numbered option.
\end{quote}

\paragraph{Scenario simulation user prompt (text-only).}
\begin{quote}
\small\ttfamily
Persona:\\
\texttt{<PERSONA\_PROMPT>}\\
\\
Scenario:\\
\texttt{<SCENARIO\_PROMPT>}\\
\\
Please think step-by-step and choose one numbered option (a number in Actions).
\end{quote}

\paragraph{Scenario simulation (with image input).}
When a scenario includes an image, we supply the same text prompt plus an associated image input \texttt{<SCENARIO\_IMAGE>}.

\paragraph{Persona prompt template (\texttt{<PERSONA\_PROMPT>}).}
\begin{quote}
\small\ttfamily
\textbf{Optional Persona Description:}\\
\texttt{<PERSONA\_DESCRIPTION>}\\
\\
\textbf{Demographic Data:}\\
sex: \texttt{<SEX>}\\
age: \texttt{<AGE>}\\
education: \texttt{<EDUCATION>}\\
income: \texttt{<INCOME>}
\end{quote}

\paragraph{Scenario prompt template (\texttt{<SCENARIO\_PROMPT>}).}
\begin{quote}
\small\ttfamily
\texttt{<SCENARIO\_DESCRIPTION>}\\
Actions:\\
1. \texttt{<ACTION\_1>}\\
2. \texttt{<ACTION\_2>}\\
\ldots
\end{quote}

\paragraph{Theory-informed prompting (optional).}
For theory-grounded conditions, we append the following instruction block:
\begin{quote}
\small\ttfamily
Respond as a typical person with the given persona would --- someone who weighs everyday trade-offs rather than acting perfectly rational or blindly trusting. People usually make quick, intuitive judgments that balance benefit, trust, effort, and risk. If a request feels clear, familiar, and low-effort, it’s normal to go along. If it feels vague, unnecessary, or slightly off, hesitation or refusal is just as natural.\\
You care about privacy and security, but you also value convenience and social normalcy. Most people don’t decide based on ideals --- they decide based on what feels reasonable right now. Sometimes they cooperate because it seems safe and routine; other times they skip or delay because the benefit isn’t obvious or the timing isn’t right.\\
Let that mix of practicality, mild caution, and situational trust guide your answer. Don’t assume participation is automatically good, or that caution always means refusal --- choose what fits how a real person like you would naturally react in that moment.
\end{quote}

\subsection{Behavior Scenario Prompts}
\label{app:simulation_prompts:phb}

For behavioral tests, we instantiate \texttt{<SCENARIO\_PROMPT>} using the scenario context and a discrete action list. For scenarios with images, we additionally provide the corresponding interface screenshot.

\paragraph{Privacy Premium.}
\begin{quote}
\small\ttfamily
\textbf{Without Privacy Cues:}\\
You have entered 'Duracell AA batteries 8-pack' into the Finder search bar and hit Search. The result page is shown in the image. Determine which action you will take.\\
Actions:\\
1. Buy from the first listing\\
2. Buy from the second listing\\
3. Buy from the third listing\\
4. Buy from the fourth listing\\
5. Spend some time to click through view each merchant's privacy policy\\

\textbf{With Privacy Cues:}\\
You have entered 'Duracell AA batteries 8-pack' into the Finder search bar and hit Search. The result page is shown in the image. Each listing includes a row of Privacy Report squares next to the title (more green box means higher privacy rating). Determine which action you will take.\\
Actions:\\
1. Buy from the first listing\\
2. Buy from the second listing\\
3. Buy from the third listing\\
4. Buy from the fourth listing\\
\end{quote}
\begin{figure}[H]
\centering
\includegraphics[width=0.95\linewidth]{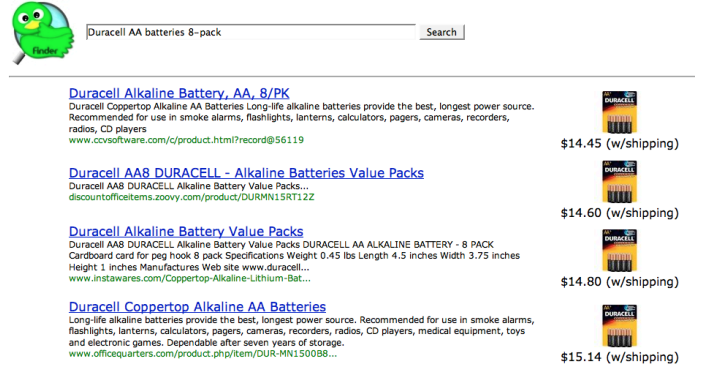}
\caption{Interface screenshot for Privacy Premium (Without Privacy Cues).}
\label{fig:phb-privacy-premium-without-cues}
\end{figure}
\begin{figure}[H]
\centering
\includegraphics[width=0.95\linewidth]{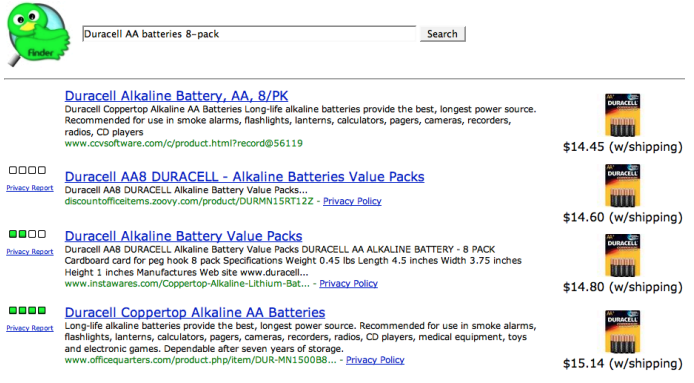}
\caption{Interface screenshot for Privacy Premium (With Privacy Cues).}
\label{fig:phb-privacy-premium-with-cues}
\end{figure}

\paragraph{Confidentiality Assurance.}
\begin{quote}
\small\ttfamily
\textbf{Without assurances:}\\
You receive an invitation to participate in a 'Citizens' Survey. It includes a general survey description suggesting a neutral, non threatening survey topic. Decide if you are willing to participate in the survey.\\
Actions:\\
1. Willing to attend\\
2. Unwilling to attend\\

\textbf{With assurances:}\\
You receive an invitation to participate in a 'Citizens' Survey. It includes a general survey description suggesting a neutral, non threatening survey topic. It also includes an elaborate confidentiality assurance and a one-page description of how the confidentiality of the data would be safeguarded. Decide if you are willing to participate in the survey.\\
Actions:\\
1. Willing to attend\\
2. Unwilling to attend\\
\end{quote}

\paragraph{Gift Card Anonymity.}
\begin{quote}
\small\ttfamily
\textbf{Having Anonymous Card:}\\
After completing a short survey at a mall kiosk, you are handed a \$10 VISA gift card labeled 'Anonymous - purchases cannot be linked to your name.' The researcher now offers: 'You can keep this \$10 anonymous card or switch to a \$12 card that will link purchases to your name.' What do you do?\\
Actions:\\
1. Keep the \$10 anonymous card\\
2. Switch to the \$12 trackable card\\

\textbf{Having Trackable Card:}\\
After completing a short survey at a mall kiosk, you are handed a \$12 VISA gift card labeled 'Trackable - purchases will be linked to your name.' The researcher now offers: 'You can keep this \$12 trackable card or switch to a \$10 anonymous card that will not be linked to your name.' What do you do?\\
Actions:\\
1. Keep the \$12 trackable card\\
2. Switch to the \$10 anonymous card\\
\end{quote}

\paragraph{Interface Formality.}
\begin{quote}
\small\ttfamily
\textbf{Professional Interface:}\\
Based on the web interface in the image, decide how you will respond. Assume you did commit everything asked.\\
Actions:\\
1. Answer questions honestly\\
2. Skip the question\\
3. Provide false answers\\

\textbf{Casual Interface:}\\
Based on the web interface in the image, decide how you will respond. Assume you did commit everything asked.\\
Actions:\\
1. Answer questions honestly\\
2. Skip the question\\
3. Provide false answers\\
\end{quote}

\begin{figure}[H]
\centering
\includegraphics[width=0.95\linewidth]{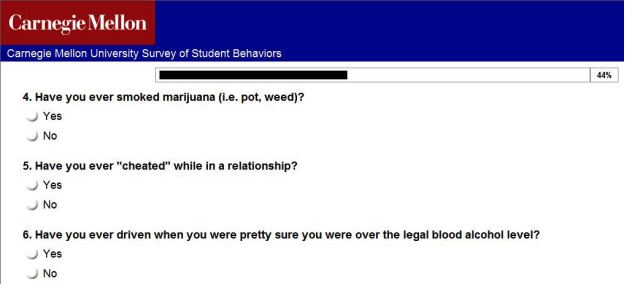}
\caption{Interface screenshot for Interface Formality (Professional Interface).}
\label{fig:phb-interface-formality-professional}
\end{figure}

\begin{figure}[H]
\centering
\includegraphics[width=0.95\linewidth]{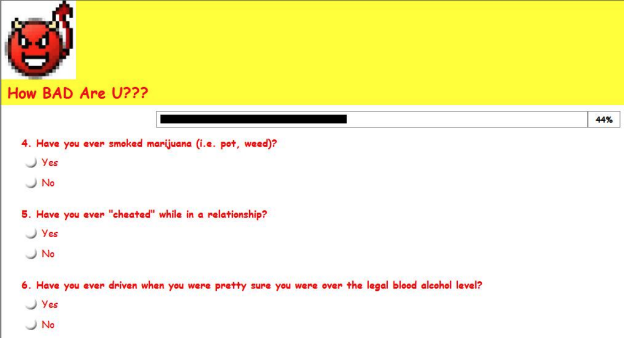}
\caption{Interface screenshot for Interface Formality (Casual Interface).}
\label{fig:phb-interface-formality-casual}
\end{figure}

\paragraph{Normative Info.}
\begin{quote}
\small\ttfamily
\textbf{Without Normative Information:}\\
You are completing an online questionnaire titled 'Study of Personal Behaviors.' The page displays one question per screen: 'Have you ever shoplifted an item under \$20?' with three radio buttons: 'Yes,' 'No,' and 'Prefer not to answer.' You did shoplift an item. Decide how you will respond.\\
Actions:\\
1. Select ``Yes''\\
2. Skip the question\\
3. Select ``No''\\

\textbf{With Low-Admission Norm:}\\
You are completing an online questionnaire titled ``Study of Personal Behaviors.'' The page displays one question per screen: ``Have you ever shoplifted an item under \$20?'' with three radio buttons: ``Yes,'' ``No,'' and ``Prefer not to answer.'' A grey banner appears beneath your response reading ``25\% of participants in an earlier survey admitted they have done this.'' You did shoplift an item. Decide how you will respond.\\
Actions:\\
1. Select ``Yes''\\
2. Skip the question\\
3. Select ``No''\\

\textbf{With High-Admission Norm:}\\
You are completing an online questionnaire titled ``Study of Personal Behaviors.'' The page displays one question per screen: ``Have you ever shoplifted an item under \$20?'' with three radio buttons: ``Yes,'' ``No,'' and ``Prefer not to answer.'' A grey banner appears beneath your response reading ``75\% of participants in an earlier survey admitted they have done this.'' You did shoplift an item. Decide how you will respond.\\
Actions:\\
1. Select ``Yes''\\
2. Skip the question\\
3. Select ``No''\\
\end{quote}

\paragraph{Unintended Audience.}

\begin{quote}
\small\ttfamily
\textbf{Without Unintended Sharing:}\\
You log in to ConnectSpace and open your Privacy Settings. Your profile visibility is currently set to 'Public,' meaning anyone on the platform can see your posts, photos, and personal info. There have been no recent notifications or incidents of unintended sharing. Decide what you want your profile visibility to be.\\
Actions:\\
1. Keep profile visibility as Public\\
2. Change profile visibility to Friends only\\

\textbf{With Unintended Sharing:}\\
You log in to ConnectSpace and see a notification: 'Alex viewed your profile.' You didn't intend to disclose your personal profile to casual acquaintances like Alex. Your Privacy Settings show 'Public.' Decide what you want your profile visibility to be.\\
Actions:\\
1. Keep profile visibility as Public\\
2. Change profile visibility to Friends only\\
\end{quote}

\paragraph{Personalized Ads.}
\begin{quote}
\small\ttfamily
\textbf{Generic Ads:}\\
You navigate to NewsHub's desktop homepage. At the top is a static banner reading 'Spring Car Sale - Save on New 2025 Sedans,' with a photo of a red sedan and no reference to your browsing history or personal data. On the right sidebar, a second ad invites you to 'Join the Summer Reading List' with a generic stack of books. There are no messages about personalization or targeting. Decide how you respond to these advertisements.\\
Actions:\\
1. Ignore the ads and continue reading or click on one of the ads\\
2. Open browser's ad settings to review personalization options\\

\textbf{Personalized Ads:}\\
You navigate to NewsHub's desktop homepage. At the top is a static banner reading ``Your Dream Sedan Awaits!'' and shows a photo of the exact red sedan you researched yesterday. It says ``Based on your searches for '2025 sedans,' get 15\% off at AutoMax.'' The sidebar ad reads ``Brooklyn JazzFest Tickets, Alex - Loved that jazz club you visited? 20\% off ends soon.'' Small text below both ads says ``Ads personalized using your browsing data.'' Decide how you respond.\\
Actions:\\
1. Ignore the ads and continue reading or click on one of the ads\\
2. Open browser's ad settings to opt out of personalization\\
\end{quote}

\paragraph{Policy Link.}
\begin{quote}
\small\ttfamily
\textbf{Without Privacy Policy:}\\
You arrive at a website's signup page (shown in image). Decide how you will respond.\\
Actions:\\
1. Provide real information, click ``Sign Up''\\
2. Provide fake or partial information, click ``Sign Up''\\

\textbf{With Privacy Policy:}\\
You arrive at a website's signup page (shown in image). Decide how you will respond.\\
Actions:\\
1. Provide real information, click ``Sign Up''\\
2. Provide fake or partial information, click ``Sign Up''\\
\end{quote}
\begin{figure}[H]
\centering
\includegraphics[width=0.95\linewidth]{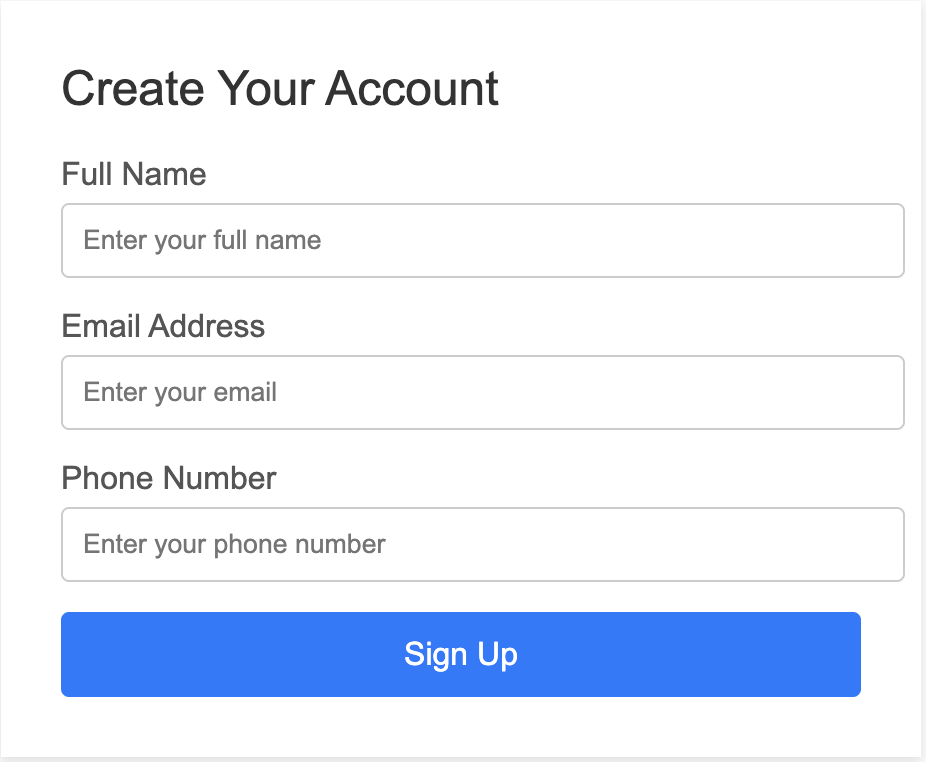}
\caption{Interface screenshot for Policy Link (Without Privacy Policy).}
\label{fig:phb-policy-link-without}
\end{figure}
\begin{figure}[H]
\centering
\includegraphics[width=0.95\linewidth]{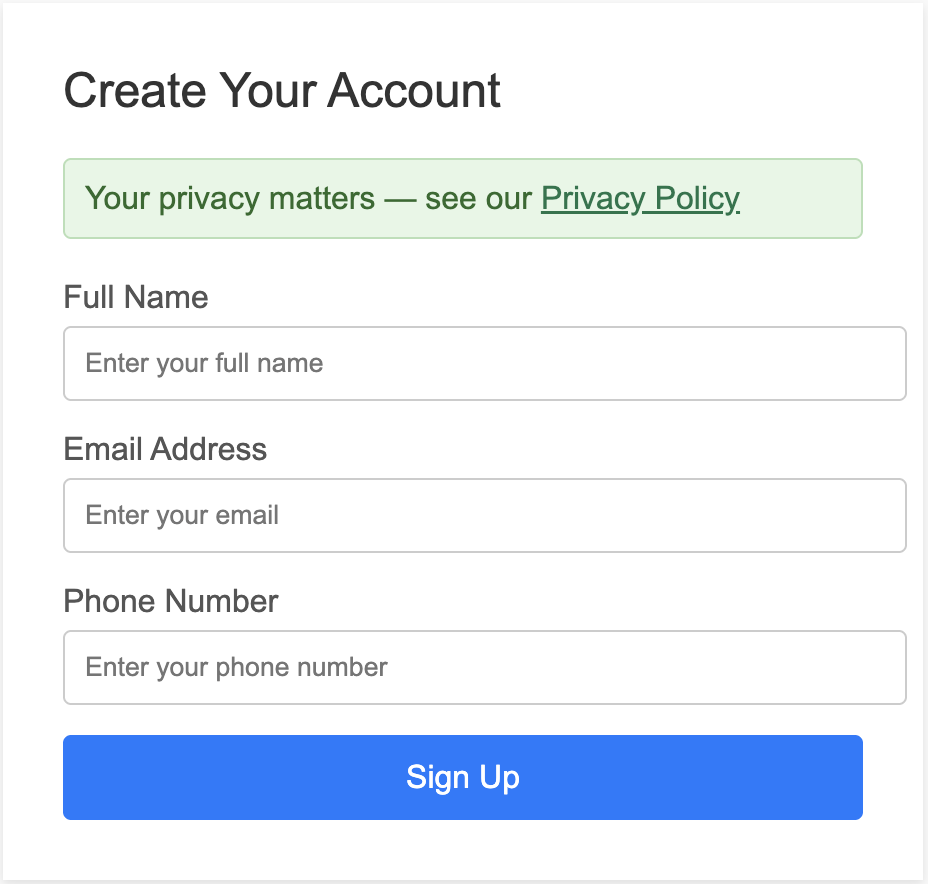}
\caption{Interface screenshot for Policy Link (With Privacy Policy).}
\label{fig:phb-policy-link-with}
\end{figure}

\paragraph{Regulatory Assurance.}
\begin{quote}
\small\ttfamily
\textbf{Without Government Regulation:}\\
You are considering signing up for a new mobile coupon service. To join, you register your mobile phone number and select categories of merchants you are interested in (e.g., restaurants, clothing stores). When you are near one of these merchants, the service will automatically send promotional information and coupons to your phone via text message. Your location will be used to target you with relevant advertisements. Would you sign up for this service?\\
Actions:\\
1. Sign up for the service\\
2. Decline the service\\

\textbf{With Government Regulation:}\\
You are considering signing up for a new mobile coupon service. To join, you register your mobile phone number and select categories of merchants you are interested in (e.g., restaurants, clothing stores). When you are near one of these merchants, the service will automatically send promotional information and coupons to your phone via text message. Your location will be used to target you with relevant advertisements. The service provider notes that your privacy is protected under the 'Privacy and Wireless Communications Protection Act.' Would you sign up for this service?\\
Actions:\\
1. Sign up for the service\\
2. Decline the service\\
\end{quote}

\paragraph{Control Paradox.}
\begin{quote}
\small\ttfamily
\textbf{Low Control:}\\
You are participating in an anonymous online survey for a research report on modern behaviors. Your answers will be used to compile the report. Please answer the following question: 'Have you ever fantasized about something you were ashamed of?'\\
Actions:\\
1. Answer the question honestly.\\
2. Decline to answer the question.\\

\textbf{High Control:}\\
You are participating in an anonymous online survey for a research report on modern behaviors. Your answers will be used to compile the report. Importantly, after you have answered all the questions, you will be able to review your individual answers and decide which, if any, will be included in the report. You have the final say.\\
\\
Please answer the following question: 'Have you ever fantasized about something you were ashamed of?'\\
Actions:\\
1. Answer the question honestly.\\
2. Decline to answer the question.\\
\end{quote}

\subsection{Parameter Settings}
\label{app:parameter_settings}

\paragraph{Temperature.}
For persona generation calls, we set \texttt{temperature}=1.0. For survey and scenario simulation calls, we set \texttt{temperature}=0.3. For \texttt{gpt-5} family models, temperature is not configurable; we therefore use the provider default.

\paragraph{Max tokens.}
We set \texttt{max\_tokens}=8192 for all API calls (persona generation and simulation).

\section{Benchmark Description}
\label{app:benchmark_description}

\begin{table*}[t]
\caption{Empirical tests comprising \textsc{SP-ABCBench}'s Attitude dimension.}
\label{tab:benchmark_empirical_tests}
\small
\begin{tabularx}{\textwidth}{@{}p{4.2cm}X@{}}
\toprule
\rowcolor{gray!20}
\textbf{Item Name} & \textbf{Description} \\
IUIPC &
Simulate the IUIPC survey, compute subscale and downstream construct scores (collection, control, awareness, trust, risk, intention), and compare each simulated score distribution to published distributions via absolute Cohen’s $d$; report the mean $|d|$ across constructs (lower is better). \\

SeBIS &
Simulate the 16-item SeBIS survey and compare each item’s response distribution to published distributions via absolute Cohen’s $d$; report the mean $|d|$ across items (lower is better). \\

SA-6 &
Simulate the SA-6 survey, compute the overall SA-6 score distribution, and compare the simulated distribution mean to the published mean via pooled-SD Cohen’s $d$; report $|d|$ (lower is better). \\

OPC &
Simulate the OPC (PUI) survey, compute subscale score distributions, and compare simulated vs. published distributions via absolute Cohen’s $d$; report the mean $|d|$ across subscales (lower is better). \\
\bottomrule
\end{tabularx}
\end{table*}

\begin{table*}[t]
\caption{Empirical tests comprising \textsc{SP-ABCBench}'s Behavior dimension.}
\label{tab:benchmark_empirical_tests}
\small
\begin{tabularx}{\textwidth}{@{}p{4.2cm}X@{}}
\toprule
\rowcolor{gray!20}
\textbf{Item Name} & \textbf{Description} \\
Privacy Premium &
In an online shopping choice, compare behavior with vs. without salient privacy-protection cues (visual privacy ratings). Score as average willingness to pay a privacy premium (higher indicates stronger privacy preference; closer to the published target indicates better alignment). \\

Confidentiality Assurance &
Given survey invitations with vs. without an elaborate confidentiality assurance, measure participation rate in each condition and score as the ratio (attendance without assurance)/(attendance with assurance); ratio $>1$ matches the published direction. \\

Gift Card Anonymity &
Under endowment with an anonymous vs. trackable gift card and an option to switch, measure (keep anonymous when offered trackable)/(switch to anonymous when endowed with trackable); ratio $>1$ matches the endowment effect and published direction and closer to the published ratio indicates better alignment. \\

Interface Formality &
Given a sensitive questionnaire presented on a professional vs. casual interface (visual), measure disclosure rate and score as (disclosure under casual)/(disclosure under professional); ratio $>1$ matches the published effect and closer to the published ratio indicates better alignment. \\

Normative Info (High vs. Low) &
For a sensitive admission question, compare disclosure rates under high-admission vs. low-admission peer norms and score as (disclosure high norm)/(disclosure low norm); ratio $>1$ matches the published direction and closer to the published ratio indicates better alignment. \\

Normative Info (High vs. Missing) &
For the same admission question, compare disclosure rates under high-admission norms vs. no normative information and score as (disclosure high norm)/(disclosure without norm); ratio $>1$ matches the published direction and closer to the published ratio indicates better alignment. \\

Unintended Audience &
On a social profile visibility setting, compare the rate of switching to friends-only when an unintended viewer is highlighted vs. when no such incident occurs; score as (switch rate with unintended audience)/(switch rate without); ratio $>1$ matches the published direction. \\

Personalized Ads &
On a news site, compare the rate of visiting privacy/ad settings after exposure to highly personalized ads vs. generic ads; score as (settings-visit rate personalized)/(settings-visit rate generic); ratio $>1$ matches the published direction. \\

Policy Link &
On a sign-up page with vs. without a visible privacy policy link (visual), measure the rate of submitting real information; score as (real-disclosure rate with policy link)/(without); ratio $>1$ matches the published direction. \\

Regulatory Assurance &
For a location-based coupon service, compare adoption with vs. without mention of a government privacy act; score as (adoption rate with regulation)/(without); ratio $>1$ matches the published direction. \\

Control Paradox &
For a sensitive disclosure prompt, compare disclosure under explicit user control over publication vs. low control; score as (disclosure low control)/(disclosure high control); ratio $>1$ matches the published direction (more disclosure under explicit control). \\

\bottomrule
\end{tabularx}
\end{table*}

\renewcommand{\arraystretch}{1.25}
\begin{table*}[t]
\caption{Empirical tests comprising \textsc{SP-ABCBench}'s Coherence dimension.}
\label{tab:benchmark_empirical_tests}
\small
\begin{tabularx}{\textwidth}{@{}p{4.2cm}X@{}}
\toprule
\rowcolor{gray!20}
\textbf{Item Name} & \textbf{Description} \\
Privacy Structural Paths &
Simulate the IUIPC survey, compute IUIPC subscales (collection, control, awareness) and downstream constructs (trust, risk, intention), fit the published structural equation model (SEM), and score coherence by mean absolute error (MAE) between simulated SEM coefficients and the published coefficients. \\

Security Intentions (DoSpeRT) &
Simulate SeBIS and DoSpeRT, compute SeBIS subscales, estimate Pearson correlations between SeBIS subscales and DoSpeRT dimensions, and score coherence by averaging absolute differences from published coefficients. \\

Security Intentions (GDMS) &
Simulate SeBIS and GDMS, compute SeBIS subscales, estimate Pearson correlations between SeBIS subscales and GDMS styles, and score coherence by averaging absolute differences from published coefficients. \\

Security Intentions (NFC) &
Simulate SeBIS and Need for Cognition (NFC), compute SeBIS subscales, estimate Pearson correlations between SeBIS subscales and NFC, and score coherence by averaging absolute differences from published coefficients. \\

Security Intentions (BIS) &
Simulate SeBIS and Barratt Impulsiveness Scale (BIS), compute SeBIS subscales, estimate Pearson correlations between SeBIS subscales and BIS, and score coherence by averaging absolute differences from published coefficients. \\

Security Intentions (CFC) &
Simulate SeBIS and Consideration for Future Consequences (CFC), compute SeBIS subscales, estimate Pearson correlations between SeBIS subscales and CFC, and score coherence by averaging absolute differences from published coefficients. \\

SA-6 Convergent (SeBIS) &
Simulate SA-6 and SeBIS, compute overall SA-6 score (mean of 6 items) and SeBIS score (mean of 16 items with instrument-specific reverse-coding), and report Spearman’s $\rho$; closer to the published $\rho$ indicates better convergent validity. \\

SA-6 Convergent (BIS) &
Simulate SA-6 and BIS, compute scale scores, and report Spearman’s $\rho$ between SA-6 and BIS; closer to the published coefficient (including sign) indicates better alignment. \\

SA-6 Convergent (GSE) &
Simulate SA-6 and General Self-Efficacy (GSE), compute scale scores, and report Spearman’s $\rho$ between SA-6 and GSE; closer to the published coefficient indicates better alignment. \\

SA-6 Convergent (SSE) &
Simulate SA-6 and Social Self-Efficacy (SSE), compute scale scores (including instrument-specific reverse-coding), and report Spearman’s $\rho$; closer to the published coefficient indicates better alignment. \\

SA-6 Demographics (Age) &
Simulate SA-6, split respondents by age (18--39 vs. 40+), compute Welch’s $t$ statistic using the paper’s contrast direction, and score coherence by closeness to the published $t$. \\

SA-6 Demographics (Gender) &
Simulate SA-6, split respondents by gender (Male vs. Female), compute Welch’s $t$ statistic using the paper’s contrast direction, and score coherence by closeness to the published $t$. \\

SA-6 Demographics (Education) &
Simulate SA-6, split respondents by education (No college vs. Attend college), compute Welch’s $t$ statistic using the paper’s contrast direction, and score coherence by closeness to the published $t$. \\

SA-6 Demographics (Income) &
Simulate SA-6, split respondents by income (Below \$25K vs. Above \$25K), compute Welch’s $t$ statistic using the paper’s contrast direction, and score coherence by closeness to the published $t$. \\

OPC Intercorrelations &
Simulate the OPC (PUI) survey, compute subscale scores (Privacy Concern, General Caution, Technical Protection), estimate inter-subscale correlations, and score coherence by MAE between simulated and published correlation coefficients. \\
\bottomrule
\end{tabularx}
\end{table*}

\cleardoublepage
\cleardoublepage
\section{Tolerance Parameter Sensitivity}
\label{app:tau_sensitivity}

To assess the robustness of our scoring procedure to the choice of tolerance parameters, we varied each $\tau$ individually while holding the others fixed at their baseline values ($\tau_1 = 2.0$ for attitude effect sizes, $\tau_2 = 0.5$ for correlations, and $\tau_3 = 2.0$ for $t$-statistics).
Specifically, we tested $\tau_1 \in \{1.5, 2.0, 2.5\}$, $\tau_2 \in \{0.25, 0.5, 0.75\}$, and $\tau_3 \in \{1.5, 2.0, 2.5\}$.
Tables~\ref{tab:tau_demographic}--\ref{tab:tau_scenario} summarize how model rankings shift under these perturbations for each persona configuration.

Across all persona configurations and tolerance variations, rankings remain highly stable.
Gemini-2.5-Flash-Lite consistently leads in Demographic persona simulations, while Gemini-3.0-Flash and GPT-4.1-Nano dominate Raw and S\&P-Primed personas.
The largest rank shifts occur in the middle of the distribution (ranks 4--8), where score differences are small, but the top and bottom performers remain consistent.
The mean absolute rank change across all $\tau$ variations is less than 1.5 positions, and Kendall's $\tau$ correlation between baseline and perturbed rankings exceeds 0.90 in all cases.
These results confirm that our findings are not artifacts of specific tolerance choices.

\begin{table*}[t]
\centering
\caption{Model rankings for \textbf{Demographic} persona under tolerance parameter variations. Baseline configuration is $\tau_1 = 2.0$, $\tau_2 = 0.5$, $\tau_3 = 2.0$ (bold columns).}
\label{tab:tau_demographic}
\small
\setlength{\tabcolsep}{4pt}
\renewcommand{\arraystretch}{1.1}
\begin{tabular}{@{}l ccc @{\hspace{1em}} ccc @{\hspace{1em}} ccc@{}}
\toprule
& \multicolumn{3}{c}{$\tau_1$ (attitude $d$)} & \multicolumn{3}{c}{$\tau_2$ (correlation)} & \multicolumn{3}{c}{$\tau_3$ ($t$-statistic)} \\
\cmidrule(lr){2-4} \cmidrule(lr){5-7} \cmidrule(lr){8-10}
Model & 1.5 & \textbf{2.0} & 2.5 & 0.25 & \textbf{0.5} & 0.75 & 1.5 & \textbf{2.0} & 2.5 \\
\midrule
Gemini-2.5-Flash-Lite & 1 & \textbf{1} & 1 & 1 & \textbf{1} & 1 & 1 & \textbf{1} & 1 \\
GPT-4.1-Mini          & 2 & \textbf{2} & 2 & 6 & \textbf{2} & 2 & 2 & \textbf{2} & 3 \\
GPT-5-Nano (Min.)     & 3 & \textbf{3} & 4 & 2 & \textbf{3} & 4 & 7 & \textbf{3} & 2 \\
Gemini-3.0-Flash      & 5 & \textbf{4} & 3 & 4 & \textbf{4} & 3 & 3 & \textbf{4} & 4 \\
GPT-5 (Min.)          & 7 & \textbf{5} & 5 & 8 & \textbf{5} & 5 & 4 & \textbf{5} & 5 \\
GPT-4.1-Nano          & 4 & \textbf{6} & 8 & 3 & \textbf{6} & 7 & 6 & \textbf{6} & 6 \\
GPT-5-Mini (Min.)     & 6 & \textbf{7} & 6 & 5 & \textbf{7} & 6 & 5 & \textbf{7} & 8 \\
GPT-5 (Med.)          & 9 & \textbf{8} & 7 & 7 & \textbf{8} & 8 & 8 & \textbf{8} & 7 \\
Llama-4-Maverick      & 8 & \textbf{9} & 10 & 10 & \textbf{9} & 9 & 9 & \textbf{9} & 9 \\
Gemini-2.5-Flash      & 10 & \textbf{10} & 9 & 9 & \textbf{10} & 10 & 10 & \textbf{10} & 10 \\
Qwen3-Next-80B        & 11 & \textbf{11} & 11 & 11 & \textbf{11} & 11 & 11 & \textbf{11} & 11 \\
GPT-4.1               & 12 & \textbf{12} & 12 & 12 & \textbf{12} & 12 & 12 & \textbf{12} & 12 \\
\bottomrule
\end{tabular}
\end{table*}

\begin{table*}[t]
\centering
\caption{Model rankings for \textbf{Raw} persona under tolerance parameter variations. Baseline configuration is $\tau_1 = 2.0$, $\tau_2 = 0.5$, $\tau_3 = 2.0$ (bold columns).}
\label{tab:tau_raw}
\small
\setlength{\tabcolsep}{4pt}
\renewcommand{\arraystretch}{1.1}
\begin{tabular}{@{}l ccc @{\hspace{1em}} ccc @{\hspace{1em}} ccc@{}}
\toprule
& \multicolumn{3}{c}{$\tau_1$ (attitude $d$)} & \multicolumn{3}{c}{$\tau_2$ (correlation)} & \multicolumn{3}{c}{$\tau_3$ ($t$-statistic)} \\
\cmidrule(lr){2-4} \cmidrule(lr){5-7} \cmidrule(lr){8-10}
Model & 1.5 & \textbf{2.0} & 2.5 & 0.25 & \textbf{0.5} & 0.75 & 1.5 & \textbf{2.0} & 2.5 \\
\midrule
GPT-4.1-Nano          & 1 & \textbf{1} & 1 & 1 & \textbf{1} & 1 & 1 & \textbf{1} & 1 \\
Gemini-3.0-Flash      & 2 & \textbf{2} & 2 & 2 & \textbf{2} & 2 & 2 & \textbf{2} & 2 \\
Gemini-2.5-Flash      & 3 & \textbf{3} & 3 & 3 & \textbf{3} & 3 & 4 & \textbf{3} & 3 \\
GPT-5-Mini (Min.)     & 4 & \textbf{4} & 4 & 4 & \textbf{4} & 4 & 3 & \textbf{4} & 4 \\
Qwen3-Next-80B        & 6 & \textbf{5} & 5 & 5 & \textbf{5} & 8 & 6 & \textbf{5} & 6 \\
GPT-4.1-Mini          & 5 & \textbf{6} & 6 & 6 & \textbf{6} & 5 & 5 & \textbf{6} & 8 \\
GPT-5 (Min.)          & 7 & \textbf{7} & 8 & 7 & \textbf{7} & 7 & 9 & \textbf{7} & 5 \\
GPT-5 (Med.)          & 8 & \textbf{8} & 7 & 9 & \textbf{8} & 6 & 7 & \textbf{8} & 7 \\
Gemini-2.5-Flash-Lite & 10 & \textbf{9} & 9 & 8 & \textbf{9} & 10 & 8 & \textbf{9} & 10 \\
Llama-4-Maverick      & 9 & \textbf{10} & 10 & 10 & \textbf{10} & 9 & 11 & \textbf{10} & 9 \\
GPT-5-Nano (Min.)     & 11 & \textbf{11} & 11 & 12 & \textbf{11} & 11 & 10 & \textbf{11} & 11 \\
GPT-4.1               & 12 & \textbf{12} & 12 & 11 & \textbf{12} & 12 & 12 & \textbf{12} & 12 \\
\bottomrule
\end{tabular}
\end{table*}

\begin{table*}[t]
\centering
\caption{Model rankings for \textbf{S\&P-Primed} persona under tolerance parameter variations. Baseline configuration is $\tau_1 = 2.0$, $\tau_2 = 0.5$, $\tau_3 = 2.0$ (bold columns).}
\label{tab:tau_sp}
\small
\setlength{\tabcolsep}{4pt}
\renewcommand{\arraystretch}{1.1}
\begin{tabular}{@{}l ccc @{\hspace{1em}} ccc @{\hspace{1em}} ccc@{}}
\toprule
& \multicolumn{3}{c}{$\tau_1$ (attitude $d$)} & \multicolumn{3}{c}{$\tau_2$ (correlation)} & \multicolumn{3}{c}{$\tau_3$ ($t$-statistic)} \\
\cmidrule(lr){2-4} \cmidrule(lr){5-7} \cmidrule(lr){8-10}
Model & 1.5 & \textbf{2.0} & 2.5 & 0.25 & \textbf{0.5} & 0.75 & 1.5 & \textbf{2.0} & 2.5 \\
\midrule
Gemini-3.0-Flash      & 1 & \textbf{1} & 1 & 1 & \textbf{1} & 1 & 1 & \textbf{1} & 1 \\
GPT-4.1-Nano          & 2 & \textbf{2} & 2 & 3 & \textbf{2} & 2 & 2 & \textbf{2} & 2 \\
Gemini-2.5-Flash      & 3 & \textbf{3} & 3 & 2 & \textbf{3} & 3 & 3 & \textbf{3} & 3 \\
GPT-4.1-Mini          & 4 & \textbf{4} & 5 & 6 & \textbf{4} & 4 & 5 & \textbf{4} & 4 \\
Gemini-2.5-Flash-Lite & 5 & \textbf{5} & 6 & 7 & \textbf{5} & 6 & 6 & \textbf{5} & 5 \\
GPT-4.1               & 6 & \textbf{6} & 4 & 4 & \textbf{6} & 5 & 4 & \textbf{6} & 6 \\
GPT-5-Mini (Min.)     & 8 & \textbf{7} & 8 & 5 & \textbf{7} & 7 & 8 & \textbf{7} & 9 \\
GPT-5-Nano (Min.)     & 7 & \textbf{8} & 9 & 9 & \textbf{8} & 8 & 7 & \textbf{8} & 7 \\
GPT-5 (Min.)          & 9 & \textbf{9} & 7 & 8 & \textbf{9} & 9 & 9 & \textbf{9} & 8 \\
Llama-4-Maverick      & 10 & \textbf{10} & 10 & 10 & \textbf{10} & 10 & 11 & \textbf{10} & 10 \\
Qwen3-Next-80B        & 11 & \textbf{11} & 11 & 11 & \textbf{11} & 11 & 10 & \textbf{11} & 11 \\
GPT-5 (Med.)          & 12 & \textbf{12} & 12 & 12 & \textbf{12} & 12 & 12 & \textbf{12} & 12 \\
\bottomrule
\end{tabular}
\end{table*}

\begin{table*}[t]
\centering
\caption{Model rankings for \textbf{Scenario-Primed} persona under tolerance parameter variations. Baseline configuration is $\tau_1 = 2.0$, $\tau_2 = 0.5$, $\tau_3 = 2.0$ (bold columns).}
\label{tab:tau_scenario}
\small
\setlength{\tabcolsep}{4pt}
\renewcommand{\arraystretch}{1.1}
\begin{tabular}{@{}l ccc @{\hspace{1em}} ccc @{\hspace{1em}} ccc@{}}
\toprule
& \multicolumn{3}{c}{$\tau_1$ (attitude $d$)} & \multicolumn{3}{c}{$\tau_2$ (correlation)} & \multicolumn{3}{c}{$\tau_3$ ($t$-statistic)} \\
\cmidrule(lr){2-4} \cmidrule(lr){5-7} \cmidrule(lr){8-10}
Model & 1.5 & \textbf{2.0} & 2.5 & 0.25 & \textbf{0.5} & 0.75 & 1.5 & \textbf{2.0} & 2.5 \\
\midrule
GPT-5-Mini (Min.)     & 2 & \textbf{1} & 2 & 1 & \textbf{1} & 1 & 2 & \textbf{1} & 2 \\
Gemini-2.5-Flash-Lite & 3 & \textbf{2} & 1 & 2 & \textbf{2} & 2 & 3 & \textbf{2} & 1 \\
GPT-5-Nano (Min.)     & 1 & \textbf{3} & 3 & 8 & \textbf{3} & 3 & 1 & \textbf{3} & 3 \\
GPT-5 (Med.)          & 5 & \textbf{4} & 4 & 6 & \textbf{4} & 4 & 5 & \textbf{4} & 5 \\
GPT-4.1-Nano          & 4 & \textbf{5} & 6 & 9 & \textbf{5} & 5 & 4 & \textbf{5} & 4 \\
GPT-5 (Min.)          & 8 & \textbf{6} & 5 & 7 & \textbf{6} & 7 & 7 & \textbf{6} & 7 \\
Gemini-2.5-Flash      & 7 & \textbf{7} & 8 & 5 & \textbf{7} & 6 & 9 & \textbf{7} & 6 \\
Qwen3-Next-80B        & 9 & \textbf{8} & 7 & 10 & \textbf{8} & 10 & 6 & \textbf{8} & 10 \\
Llama-4-Maverick      & 6 & \textbf{9} & 10 & 4 & \textbf{9} & 8 & 8 & \textbf{9} & 9 \\
GPT-4.1               & 10 & \textbf{10} & 9 & 3 & \textbf{10} & 9 & 10 & \textbf{10} & 8 \\
Gemini-3.0-Flash      & 11 & \textbf{11} & 11 & 11 & \textbf{11} & 11 & 11 & \textbf{11} & 11 \\
GPT-4.1-Mini          & 12 & \textbf{12} & 12 & 12 & \textbf{12} & 12 & 12 & \textbf{12} & 12 \\
\bottomrule
\end{tabular}
\end{table*}

\cleardoublepage
\cleardoublepage
\section{Additional Figure}
\label{app:additional_figure}

\begin{figure*}[t]
\centering
\includegraphics[width=1.0\textwidth]{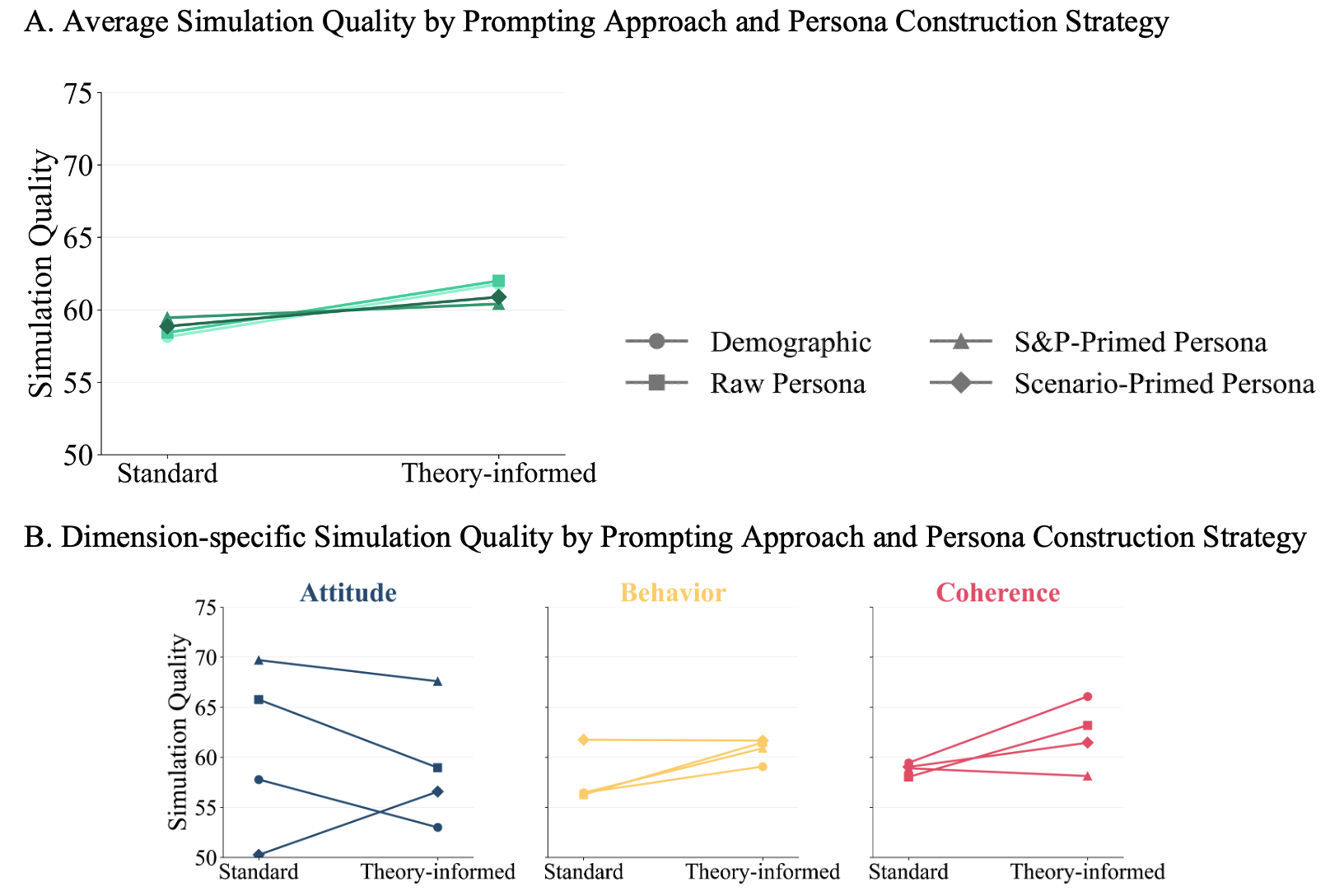}
\caption{Theory-informed prompting effects by persona construction strategy across four models. (A) Overall: all strategies improve, with Demographic showing the largest gain. (B) Dimension-specific interactions: For Behavior, all strategies improve except Scenario-Primed which shows negligible change; for Attitude, all strategies degrade except Scenario-Primed which improves substantially; for Coherence, all strategies improve except S\&P-Primed which slightly declines.}
\label{fig:tip-pcs}
\end{figure*}

\end{document}